\documentclass[12pt]{iopart}
\usepackage{graphicx}				
\usepackage{iopams}
\usepackage{url}		
\usepackage{hyperref}
\usepackage[numbers,sort&compress]{natbib}
\usepackage{hypernat}	
\usepackage[american]{babel}
\begin{document}
\selectlanguage{american}
\title{Pedestrian Traffic: on the Quickest Path}
\author{Tobias Kretz}
\address{PTV Planung Transport Verkehr AG\\
Stumpfstra{\ss}e 1\\ D-76131 Karlsruhe, Germany}
\ead{Tobias.Kretz@PTV.De}

\maketitle
\begin{abstract}
When a large group of pedestrians moves around a corner, most pedestrians do not follow the shortest path, which is to stay as close as possible to the inner wall, but try to minimize the travel time. For this they accept to move on a longer path with some distance to the corner, to avoid large densities and by this succeed in maintaining a comparatively high speed. In many models of pedestrian dynamics the basic rule of motion is often either ``move as far as possible toward the destination'' or -- reformulated -- ``of all coordinates accessible in this time step move to the one with the smallest distance to the destination´´. Atop of this rule modifications are placed to make the motion more realistic. These modifications usually focus on local behavior and neglect long-ranged effects. Compared to real pedestrians this leads to agents in a simulation valuing the shortest path a lot better than the quickest. So, in a situation as the movement of a large crowd around a corner, one needs an additional element in a model of pedestrian dynamics that makes the agents deviate from the rule of the shortest path. In this work it is shown, how this can be achieved by using a flood fill dynamic potential field method, where during the filling process the value of a field cell is not increased by 1, but by a larger value, if it is occupied by an agent. This idea may be an obvious one, however, the tricky part -- and therefore in a strict sense the contribution of this work -- is a) to minimize unrealistic artifacts, as na{\"i}ve flood fill metrics deviate considerably from the Euclidean metric and in this respect yield large errors, b) do this with limited computational effort, and c) keep agents' movement at very low densities unaltered.\\
\noindent
\end{abstract}

\tableofcontents

\section{Introduction}
\subsection{Pedestrians in the Light of Light}
Everyone knows from own experience that even when selecting the mode of travel but also as a participant in any mode of traffic minimizing the travel time is an important criterion when planning a trip. There can be other criteria against which the shortest travel time is weighed against for example monetary costs or sometimes pedestrians value staying away from large streets high. However, minimizing the travel time is almost never irrelevant. If the travel time is not minimized for a full path one can often insert intermediate destinations on a path, to make the rule of minimal travel times apply for each of the partial paths.

As a first step of modeling let's neglect all other criteria and assume for pedestrians: 
\begin{quote} \em
Pedestrians try to move from their current position to their destination on the path of shortest travel time.
\end{quote}

This bears a remarkable similarity to the formulation of Fermat's Principle for the path of light: 
\begin{quote} \em
On the way from a given point A to a given point B a ray of light takes the path of least time.\em \cite{Fermat1662}
\end{quote}

Why then is the motion of pedestrians and light so different? It's not only the kind of interaction, but a more fundamental difference: Fermat's Principle is only seemingly teleological, while in fact it is the result of laws of nature following causality. Classically this follows from Huygen's Principle \cite{Huygens1690} and quantum-mechanically from Feynman's Path Integral Formalism \cite{Feynman1965}. I.e. for certain calculations the teleological and the causal formulation are {\em equivalent}. Pedestrians on the contrary act teleologically as they have -- at least in introspective -- a free will and make plans. {\em Additionally} causal restrictions apply to their motion. If one restricts the motion of participants in an experiment to a one-dimensional motion, the focus is set to these causal restrictions. In such a case there is no more teleological element present. An experiment purely investigating this is for example reported about in \cite{Seyfried2005,Seyfried2008}. In other cases experiments and observations often do not exclusively investigate the causal restrictions but still emphasize it a lot, as there is usually a very limited set of alternatives available for any kind of planning process \cite{Fruin1971,Predtetschenski1971,Hoogendoorn2003a,Fujiyama2006,Kretz2006i,Kretz2006h,Seyfried2006,Johansson2007}.

As a result of this double influence -- causal and teleological -- and because of limited access to information of the path ahead -- pedestrians only {\em try} to minimize their time and usually only partially succeed with that. On the contrary for light there are only few clearly defined and well understood exceptions from the path of minimal time: in very intense (many photons) ray of light with very high energy (frequency) it can happen with non-negligible probability that light scatters at light \cite{Euler1936} and the ray is ``fanned out" and -- just for completeness -- gravitational lenses can produce images with light moving on a path of (locally) maximal time \cite{Blandford1986}. The case of gravitational lenses is not of much interest here. But light-light scattering is of some interest as with it the applicability of Fermat's Principle vanishes for light just for a similar reason why Fermat's Principle cannot be used to do calculations of pedestrian dynamics. Within a normal course of life beam of light photons typically do not interact as the photon-photon scattering probability (its cross section) is by far too small. \footnote{The reason for this being that charge conservation does not allow a tree-level process. So the scattering in lowest order involves two electron propagators which in the end lead to the electron mass to the eighth in the denominator. Furthermore the four vertices bring in an $\alpha^4$, which decreases the scattering probability further. Only if energy and density of the photons is large enough these factors can be balanced to non-negligible overall scattering probabilities.}

As long as the photons within a ray of light do not scatter, Fermat's Principle expressed in a formula says that the travel time
\begin{equation}
\tau = \int_{Path P: \vec{x}=\vec{A}}^{\vec{x}=\vec{B}} \frac{d|\vec{x}|}{v(\vec{x})}
\end{equation}
is minimal for the path $P$ ``chosen" by the ray of light. $v(\vec{x})=c/n(\vec{x})$ is the speed of light as the case may be within some material with some refractivity $n(\vec{x})$. From knowing the field of speeds $|\vec{v}(\vec{x})|$ one could calculate the field $T(\vec{x})$ of minimal travel times to the destination $\vec{B}$. For each position the direction of the ray of light then would be equal to the one of the gradient $\nabla T(\vec{x})$, as Fermat's principle compulsorily also holds for each partial path.

Classical optics is simpler than high energy quantum optics or the movement of pedestrians as $v(\vec{x})$ only depends on the material and its refractivity at position $\vec{x}$ and not the light itself or to be more precise the spatial-temporal distribution of photons. For pedestrians $\vec{v}(\vec{x},t)$ heavily depends on the local distribution of other pedestrians at a certain time. Put mathematically: the velocity field $\vec{v}(\vec{x},t)$ -- or even more general: the field of velocity distributions -- is not an input but a result of the calculation of pedestrian dynamics and those space-time regions are of specific interest, where the speed is significantly lower than the desired speeds: ``jams". Obviously one is faced with a highly non-linear phenomenon.

Having realized that the most important input factor is not available in pedestrian dynamics, the first and radical step is to assume that the agents do not influence each other at all. This is equivalent to allowing two agents to take the same place at the same time, which of course is not realistic and would result in all agents walking the same spatially shortest path regardless of the density on that path. So, the minimal mutual influence of agents is a strong, short-ranged, repulsive influence, often simply a hard-core exclusion, respectively a hard-core repulsion. 

As apart from this exclusion principle at this stage no additional interaction between agents exists, the basic direction of motion then is determined from the gradient of the field of spatial distances to the destination: $\nabla S(\vec{x})$. The element of a model that incorporates the influence of $\nabla S(\vec{x})$ mathematically often has the same structure as the other elements of the model. Its special status is then reflected in the typically most important role the parameters configuring the strength of the influence of $\nabla S(\vec{x})$ have.

Then as a third step the actual modeling process sets in, where all the available models differ from each other and deviations from the basic direction of motion are calculated. Normally at least one of these additional elements is meant to compensate for the agents following $\nabla S(\vec{x})$ too strictly, i.e. walking too close to the spatially shortest path and therefore producing unrealistic jams. At this point the modeler is faced with a problem: In many situations this element of the model would have to incorporate quite long-ranged interactions to make the movement of agents realistic. But long-ranged interactions often are computationally costly. In a majority of cases modelers decide in favor of quick computation and against long-ranged interactions, accepting the formation of unrealistic jams in certain situations \cite{Rogsch2007}, as the agents esteem the spatially shortest path too much compared to the quickest path.  For a notable exception, where a minimized travel time is considered in an integrated way, see \cite{Hoogendoorn2004a,Hoogendoorn2004b}.

This work investigates the question, if the field (aka ``potential") $S(\vec{x})$ can be altered with some method to some field $\hat{S}(\vec{x},t)$ (here a time dependence must be introduced, because $\hat{S}(\vec{x},t)$ necessarily depends on the distribution of agents), such that the basic direction of motion in general is more similar to the case as if it had been calculated from the unknown $T(\vec{x},t)$.

It is assumed that the pedestrians to be simulated have fairly good knowledge of their environment. Contrary to an ansatz like in \cite{Teknomo2007}, where agents' poor knowledge of the environment is intended to be handled in the operational model, this aspect is spared here, as it is believed that this should be handled on the tactical or strategical level of scenario modeling: an agent has perfect knowledge of the spatial structure relevant for the path between his current origin-destination relation. If this was not the case, the scenario modeler would have to split up the origin-destination relation into smaller pieces until the condition holds. Incomplete information then would be reflected in the routing decision at an origin with multiple potential destinations attached and not all over the path between origin and destination. For a description of the different levels of modeling see for example \cite{Hoogendoorn2002b}.

\subsection{The Use of Potentials in Pedestrian Dynamics and Robotics}
Potential fields have long been used in robotics \cite{Khatib1986,Latombe1991} (let's call them $U(\vec{x})$ here) and the simulation of pedestrians \cite{Schadschneider2001b,Kirchner2002b,Kirchner2002,Nishinari2004,Kretz2006f,Kretz2007a,Schadschneider2009} (let's call them $S(\vec{x})$ here) to control robots and agents to evade obstacles and move toward a destination. The difference is that in robotics it is usually assumed that an \emph{autonomous robot} knows about its destination coordinate but has no knowledge of the position of obstacles except for those which it ``sees''. For pedestrians on the contrary it is typically assumed that they have at least some knowledge on the whole path, the positions of static obstacles and the detours they have to walk compared to linear distance, even if they actually only see a small fraction of the whole path. This has consequences for the calculation and use of the potential. In robotics the \emph{artificial potential} $U$at position $\vec{x}$ from a destination at $\vec{x}_d$ was originally \cite{Khatib1986} calculated as
\begin{eqnarray}
U_{artificial}(\vec{x}) & = & U_{\vec{x}_d}(\vec{x}) + U_{obstacles}(\vec{x})\\
U_{\vec{x}_d}(\vec{x}) & = & \frac{1}{2} k_d (\vec{x} - \vec{x}_d)^2 \label{eq:U}
\end{eqnarray}
In such a potential local minima can occur and a robot has to be equipped with the ability to realize it is in a minimum and how to get out of it. Much different from this is the basic assumption in many models of pedestrian motion that pedestrians have a good global knowledge of the exact Euclidean distances and know about the shortest path between their current position and their destination under consideration of all obstacles. This assumption can be unrealistic for very complex geometries like huge mazes, but for many situations it comes close to reality. In this contribution ``potential'' is used in the sense of pedestrian dynamics.

The double usage of the word ``potential'' has repeatedly led to confusion. For example in \cite{Teknomo2007} a number of models have been attributed to use robotics-like potentials, although they do not. In contrast the ``sink propagation value'' method proposed there is very similar to the static floor field used in \cite{Schadschneider2002a} rendered more precisely in \cite{Nishinari2004}, and \cite{Kretz2006d}.

In publications of the Cologne-Tokyo school of pedestrian dynamics modeling \cite{Schadschneider2001b,Nishinari2004} the potential is called ``static floor field''. Similar to \cite{Teknomo2007} the values of the static floor field decrease with increasing distance from the destination, linearly reflecting the attractiveness of a cell. In \cite{Kretz2008c} a computer science perspective has been adopted: the distance to the destination is a relevant input value, so a look up table of distances is written to memory. The difference to the static floor field is trivial, but to not get messed up and miss a sign, the latter field has been called ``distance potential field''.

There are many methods to calculate a potential $S(\vec{x})$ that consider obstacles, when calculating distances. Flood fill methods, Dijkstra's algorithm on a visibility graph \cite{Dijkstra1959,deBerg1997,Nishinari2004,Kretz2007a}, the fast marching method \cite{Kimmel1998} and ray casting \cite{Kretz2008c}. These methods have been compared in \cite{Kretz2008c} or used and applicated in \cite{VISSIM2008} and it has been concluded that in large real world applications only flood fill methods are quick enough to be calculated each time step or each $N$ time steps, with $N$ chosen such that the calculation is done for example each simulation second. Flood fill methods, however, yield large errors compared to true Euclidean distances \cite{Kretz2008c}.

\subsection{Floor Field Models of Pedestrian Dynamics}
In floor field models \cite{Schadschneider2001b,Schadschneider2002a,Kirchner2002b,Kirchner2002,Nishinari2004,Kretz2006d,Kretz2006f,Kretz2007a,Kretz2008f,Henein2008} agents move on a discrete lattice from cell to cell. Before an agent moves, a probability is assigned to each cell he could reach and then a destination cell is selected randomly according to these probabilities. Most influences on an agent's motion are modeled as partial probabilities following an exponential function
\begin{equation}
p_X=e^{k_X\cdot func(X)},
\end{equation}
where $k_X$ is a coupling constant, denoting the strength of the influence. The partial probabilities are multiplied to form the full probability that a cell is chosen as destination
\begin{equation}
p=N\prod_Xp_X.
\end{equation}
$N$ normalizes the probability such that the sum over all accessible cells equals 1.
For this work it is sufficient to assume that all previously introduced influences, except for the influence of the static floor field (resp. distance potential field) -- containing in each cell the information on the distance of that cell $(x,y)$ to the destination -- are switched off: $k_X=0$, except for $k_S$,
\begin{equation}
p_{xy}=Ne^{-k_SS_{xy}} \label{eq:S}
\end{equation}

\section{Altering the Distance Potential Field -- A Quick Method as Perturbation of an Exact One}
As has been stated, there are methods that are sufficiently precise in the calculation of distances, such that no unrealistic looking artifacts in the motion of the agents can be observed in a simulation. But these methods need painfully large times for their calculation in large scenarios, if they are executed each simulation second. And there are methods, which are in principle quick enough to calculate the potentials each time step, but these methods often make the agents move on strange trajectories, as they are not very precise with regard to Euclidean metric.

The idea is, to calculate an exact and precise ($S=S(\vec{x})$) potential and a potential using a quick method ($S_{dyn}=S_{dyn}(\vec{x},t=0)$) potential in advance, i.e. before any agent has entered the simulation. And then calculate each time step -- dynamically -- with the quick method a potential under consideration of the agents: ($S^t_{dyn}=S_{dyn}(\vec{x},t)$). The probability to select a cell as destination is then calculated as
\begin{equation}
p=Ne^{-k_S(S+S^t_{dyn}-S^0_{dyn})}  \label{eq:S1}
\end{equation}
For sake of clarity the x and y indices of equation (\ref{eq:S}) have been dropped. With this method the imprecise but quick method becomes a perturbation of the exact method. If there is only one agent in the scenario, he is only minorly affected, as all perturbation of the field is ``behind'' him, i.e. further away from the destination than he himself is. As only differences between $S^t_{dyn}$ and $S^0_{dyn}$ influence the motion, one can hope that this difference also cancels out a large share of the errors of the quick method. Introducing a new coupling constant for the new effect, one can generalize equation (\ref{eq:S1}) to
\begin{equation}
p=Ne^{-k_SS-k^,_S(S^t_{dyn}-S^0_{dyn})}=Ne^{-k_SS}e^{-k^,_S(S^t_{dyn}-S^0_{dyn})}= Np_Sp^,_S \label{eq:S2}
\end{equation}
In this way an entirely new influence has been created, fully independent from the influence of the static floor field. It cannot be called dynamic floor field, as this is the name for the virtual trace \cite{Schadschneider2001b,Nishinari2004}, so it is referred to as ``dynamic distance potential field'' here.

\section{The Quick Method: Flood Filling}
It is not only that the exact methods take too long for their calculation, also only a subset of them would be able to consider the agents in one way or another meaningfully as a modification. This is different with flood fill methods. In the process of flooding the scenery, one simply adds 1, if the cell is not occupied and a value $s_{add}>1$, if it is occupied by an agent. Basically there are two simple flood fill methods: the first acts in the von Neumann neighborhood. It is called the Manhattan metric and is the $p=1$ vector norm. The other one acts on the Moore neighborhood. It is called the Chebyshev metric and is the $p\rightarrow\infty$ vector norm. A combination of these is variant 1 (V1) as introduced in \cite{Kretz2008c}, where the maximum error is limited to smaller values than in the two basic methods on cost of having to do two flood fills and calculate a square root for each cell.

The method proposed introduces two new parameters to the model: the coupling $k_{Sdyn},$ and the summand $s_{add}$ in case a cell is occupied.

\subsection{The Role of Parameter $s_{add}$}
While the effect of $k_{Sdyn}$ is obvious and just the same as the effect of all other coupling constants in a floor field model, the situation is a bit different for $s_{add}$. A large $s_{add}$ does not in general increase the strength of the effect, this is only true for those cells on which agents are located. For all other cells $s_{add}$ determines the size a block of occupied cells needs to have that the flood partially flows through the block and not around. This does not only depend on $s_{add}$ but also on the shape of the block: there's only an effect, if from the potential's wave front perspective the block is wider than deep. If three agents are standing next to each other, all on the same value of the distance potential field, the value behind the central one is 3, if $s_{add}=2$ and 4 for any other value of $s_{add}$ (assumed $s_{add} \in \mathcal{N} $ and Manhattan metric is used). If the solid block of occupied cells is deeper than wide, the flood always has to flow around the block. In other words: for large values of $s_{add}$ blocks of occupied cells act as if they were obstacles and the flood has to flow around them during the calculation of the potential. Only if such blocks would bar the flood's way to some part of the area or if $s_{add}$ is small and the blocks wider than deep the flood can also flow through these blocks. The former case can be interpreted as a block being infinitely wide (from border to border). 

It's also blocks of agents ranging from border to border -- not the idea that agents are in some way ``softer" than obstacles -- that are the reason, why agents can not be treated like obstacles, i.e. that $s_{add}$ is set to infinity, respectively that the dynamic distance potential field may not flood ``over" agents at all. Otherwise the flooding algorithm could not reach parts of the area, if somewhere there is a chain of agents without a gap from one border to another. 

So there are two interpretations of $s_{add}$: if there is no line of agents ranging from one border to another $s_{add}$ is an estimation for the time to walk around the block, the bigger the block, the larger the time. If the block does range from one border to another $s_{add}$ is a measure for the speed reduction or time loss. The length of a jam ranging from border to border times $s_{add}-1$ then determines the length of an accepted detour.

\subsection{Examples}
\begin{figure}[htbp]
  \center
	\includegraphics[width=0.3\textwidth]{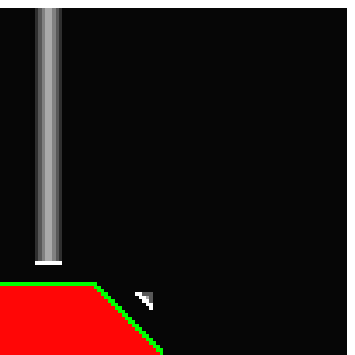} \hspace{6pt}
	\includegraphics[width=0.3\textwidth]{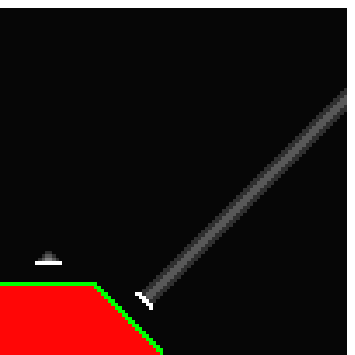} \hspace{6pt}
	\includegraphics[width=0.3\textwidth]{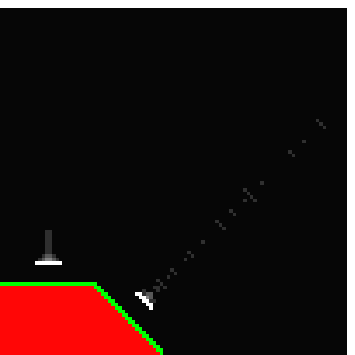}
	\caption{$S_{dyn}^t - S_{dyn}^0$ ($s_{add}=10$; from left: Manhattan, Chebyshev, V1) with 17 agents placed on two lines at the brightest white spots (exits are shown green, walls red). The difference of the time dependent and the empty potential is the larger, the brighter a spot is.}
	\label{fig:Pot1}
\end{figure}

\begin{figure}[htbp]
  \center
	\includegraphics[width=0.3\textwidth]{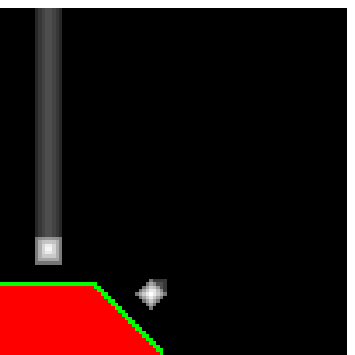} \hspace{6pt}
	\includegraphics[width=0.3\textwidth]{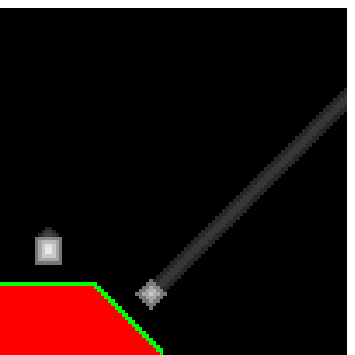} \hspace{6pt}
	\includegraphics[width=0.3\textwidth]{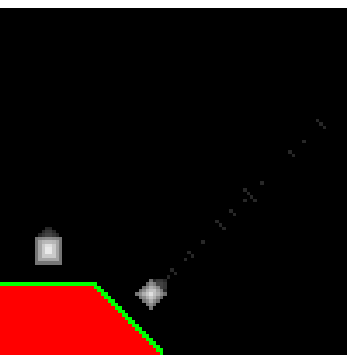}
	\caption{$S_{dyn}^t - S_{dyn}^0$ ($s_{add}=10$; from left: Manhattan, Chebyshev, V1) with 105 agents placed at the bright white squares. The difference of the time dependent and the empty potential is the larger, the brighter a spot is.}
	\label{fig:Pot2}
\end{figure}

\begin{figure}[htbp]
  \center
	\includegraphics[width=0.3\textwidth]{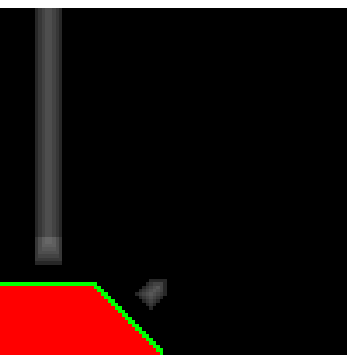} \hspace{6pt}
	\includegraphics[width=0.3\textwidth]{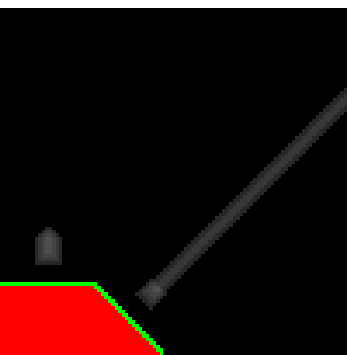} \hspace{6pt}
	\includegraphics[width=0.3\textwidth]{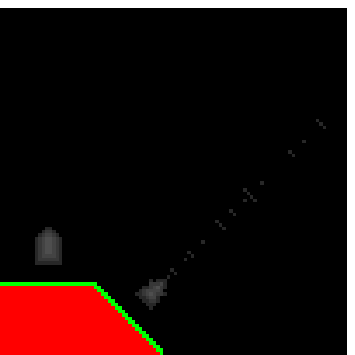}
	\caption{$S_{dyn}^t - S_{dyn}^0$ ($s_{add}=2$; from left: Manhattan, Chebyshev, V1) with 105 agents placed on two squares. The difference of the time dependent and the empty potential is the larger, the brighter a spot is.}
	\label{fig:Pot3}
\end{figure}

Figures \ref{fig:Pot1} to \ref{fig:Pot3} show the impact of a group of agents on $S_{dyn}^t$ by showing $S_{dyn}^t - S_{dyn}^0$. It can be seen clearly, that with Manhattan metric the impact spreads infinitely in vertical (and horizontal) direction and with Chebyshev metric in diagonal direction. Only with V1 the impact range is finite in any case. The seemingly strange pattern into diagonal direction with V1 metric is a result of rounding effects: instead of all cells' value approaching to zero, with increasing distance less and less cells have a value 1.

\section{Comparison to a Static Method}
The main problem, why this method is introduced is that without some method to prevent agents to crowd too densely at the inner side of the corner, the flow is reduced unrealistically at the corner, as only a tiny space at the corner is used by the agents. It has been shown
\cite{Nishinari2004,Kretz2007a} that this can be achieved partially by a simpler, static method, where the probability that a cell is chosen as destination is reduced, if its distance $W$ to a wall is smaller than some distance $W_{max}$:
\begin{equation}
p_W=Ne^{-k_W\max(0,W_{max}-W)}
\end{equation}
However, this method either assumes that large amounts of agents will flow around a corner. Then $W_{max}$ has to be large to distribute the agents on a wide variety of radii from 0 to $W_{max}$ around the corner. But then, if a small number of agents arrives, they will keep an unrealistically large distance to the corner. Or, if one expects small numbers, $W_{max}$ has to be small, and for large groups the effect will be too small.

The stadium example of subsection \ref{sec:Stadium} was calculated with the static method to make a comparison; see the discussion there.

\section{Transferability}
The method was introduced above using the F.A.S.T. model \cite{Kretz2006f,Kretz2007a,Kretz2008f} as an example, but its application is by no means limited to this or any other floor field model. It can for example also be used in a continuous space model like the social force model \cite{Helbing1995,Helbing2000b,Helbing2002,Werner2003,Johansson2007,VISSIM2008}. For a model that is continuous in space and not cell-based, one has to integrate the cell concept to the else continuous concept and therefore answer questions like ``Should agents always be assigned to one cell, or can there be parts of agents on cells?'', ``What to do, if more than one agent occupies a cell?'', or ``How large should the cells be chosen?''. Probably details of the answer will only affect details of the results. Finally, $\Delta S^t_{dyn} = S^t_{dyn}-S_{dyn}$ can be used as a modification in the method, where the direction of the desired motion is determined.

\section{Discussion: Examples of the Influence of the Method}
\subsection{The Influence on Free Speed}
The influence on free speed of this method is model specific. For the tests with the F.A.S.T. model the following settings were made: $k_S=1.0$, $k_{Sdyn}=10.0$, $k_{any other} = 0.0$, and $s_{add}=10$. All simulations were repeated 100 times with identical settings. One single agent was placed 250 cells away from his destination. The room was 400 X 400 cells wide, such that border effects could if at all only play an insignificant role. The following results for the number of rounds to reach the destination were obtained (format: average $\pm$ std (min, max)):\\
\begin{table}[htbp]
\center
\begin{tabular}[htbp]{c|cccc}
$v_{max}$& none            & Manhattan       & Chebyshev      & V1               \\ \hline
   1     & 437.2 $\pm$ 21.6& 368.5 $\pm$ 14.0& 400.3 $\pm$ 19.9& 401.0 $\pm$ 21.2 \\
         &       (388, 492)&       (326, 397)&       (346, 459)& (351, 454)       \\
   5     & 61.32 $\pm$ 2.18& 60.87 $\pm$ 2.56& 60.97 $\pm$ 2.71& 61.15 $\pm$ 2.33 \\
         &       (57, 69)  &       (56, 67)  &       (55, 69)  & (57, 67)         \\
\end{tabular}
\end{table}

So there is a some influence at $v_{max} = 1$ and no recognizable influence at $v_{max} = 5$ and for $v_{max} = 1$ the influence is stronger, when the Manhattan metric is used. This is due to the motion along the discretization axis. 

These results can be obtained analytically. If one sets the smallest value of the static distance potential field and the dynamic distance potential field to 0, it will always look like this, if the agent is meant to move downward:\\
Exact static distance potential field:\\
\begin{table}[htbp]
\center
\begin{tabular}[h]{|c|c|c|} \hline
2&2&2\\ \hline
1&1&1\\ \hline
0&0&0\\ \hline
\end{tabular}
\end{table}

The Manhattan and Chebyshev dynamic distance potential fields are shown in table \ref{tab:fields}.\\
\begin{table}[htbp]
\center
\begin{tabular}[h]{|c|c|c|} \hline
2&3&2\\ \hline
1&10&1\\ \hline
0&0&0\\ \hline
\end{tabular}
\hspace{12pt}
\begin{tabular}[h]{|c|c|c|} \hline
2&2&2\\ \hline
1&10&1\\ \hline
0&0&0\\ \hline
\end{tabular}
\caption{Manhattan and Chebyshev dynamic distance potential fields.}
\label{tab:fields}
\end{table}

This leads to exponents in the probabilities (with $k_S=1.0$ and $k_{Sdyn}=10.0$) as shown in table \ref{tab:Exponents}\\
\begin{table}[htbp]
\center
\begin{tabular}[h]{|c|c|c|} \hline
-2&-12&-2\\ \hline
-1&-91&-1\\ \hline
0&0&0\\ \hline
\end{tabular}
\hspace{12pt}
\begin{tabular}[h]{|c|c|c|} \hline
-2&-2&-2\\ \hline
-1&-91&-1\\ \hline
0&0&0\\ \hline
\end{tabular}
\caption{Exponents of probabilities for Manhattan and Chebyshev metric.}
\label{tab:Exponents}
\end{table}

and therefore the following expectation values for the time to reach the destination:
\small
\begin{eqnarray*}
\bar{T}_{none}(250) &=& 250 / \frac{1\cdot3e^0+0\cdot3e^{-1}+(-1)\cdot3e^{-2}}{3e^0+3e^{-1}+3e^{-2}} \approx 434.6\\
\bar{T}_{Manhattan}(250) &=& 250 / \frac{1\cdot3e^0+0(2e^{-1}+1\cdot e^{-91})+(-1)(2e^{-2}+e^{-12})}{3e^0+2e^{-1}+e^{-91}+2e^{-2}+e^{-12}}  \approx 366.9\\
\bar{T}_{Chebyshev}(250) &=& 250 / \frac{1\cdot3e^0+0(2e^{-1}+1e^{-91})+(-1)\cdot3e^{-2}}{3e^0+2e^{-1}+e^{-91}+3e^{-2}}  \approx 399.2
\end{eqnarray*} 
\normalsize

If one does the same with $v_{max}=5$ one sees that all terms that bring in differences to the case without dynamic distance potential are small compared to those terms that are there in any case, which makes all changes vanish. Graphically the reason is that the neighborhood for $v_{max}=5$ is wider, but there is still at maximum one cell influenced by the dynamic distance potential. This in turn implies that if there are a few agents in the neighborhood of an agent, there can again be an accelerating effect at all maximum speeds. But all of this depends on the ratio $k_{Sdyn} / k_S$. If it is small this effect will also be small and it must not be forgotten that for this investigation it was set to an absurdly large value of 10.

\subsection{A Crowd Moving Around a Corner}
This is the elementary situation for which the method was introduced, i.e. where the biggest problems occur, if the basic direction is calculated from $S(\vec{x})$.
\begin{figure}[htbp]
  \center
	\includegraphics[width=0.36\textwidth]{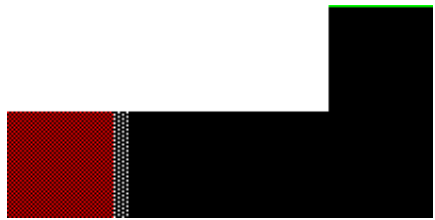}
	\caption{Starting position of 1250 agents (red) to move along a corridor (black) around a corner to a destination (green). The walls are white. The inflow was restricted by a number of columns.}
	\label{fig:Ecke1}
\end{figure}
As shown in figure \ref{fig:Ecke1} 1250 agents had to move around a corner. The maximum speed was set to 5 for all. All simulation runs were repeated 100 times. Without dynamic distance potential field the process took between 308 and 319 rounds, at an average of $314.1\pm 2.4$ rounds. The results for the various metrics and different parameters is shown in \ref{tab:Ecke1}.

\begin{table}[htbp]
\center
\small
\begin{tabular}[htbp]{c|ccc}
$k_{Sdyn}$ / $s_{add}$ &Manhattan       & Chebyshev                & V1\\ \hline
1.0 / 2     & 224.5 $\pm$ 2.2 (219, 231)& 267.8 $\pm$ 1.8 (263, 267)& 249.5 $\pm$ 1.9 (246, 255) \\
3.0 / 2     & 191.4 $\pm$ 2.0 (186, 197)& 242.1 $\pm$ 1.9 (237, 247)& 214.1 $\pm$ 1.8 (211, 220) \\
1.0 / 4     & 185.0 $\pm$ 1.7 (182, 190)& 246.7 $\pm$ 1.8 (243, 252)& 207.3 $\pm$ 2.0 (203, 215) \\
3.0 / 4     & 178.7 $\pm$ 1.8 (174, 183)& 235.7 $\pm$ 2.0 (231, 241)& 181.7 $\pm$ 1.5 (178, 185) \\
1.0 / 10    & 176.0 $\pm$ 2.1 (171, 181)& 243.9 $\pm$ 1.7 (240, 249)& 184.2 $\pm$ 1.7 (181, 190) \\ 
10.0 / 10   & 190.3 $\pm$ 2.5 (185, 198)& 240.5 $\pm$ 2.5 (234, 247)& 196.4 $\pm$ 2.2 (190, 205) \\
\end{tabular}
\caption{Number of rounds until the last agent has walked around the corner and left the simulation}
\label{tab:Ecke1}
\normalsize
\end{table}

\begin{figure}[htbp]
  \center
	\includegraphics[width=0.30\textwidth]{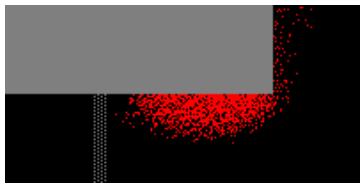}
	\caption{Situation after 90 seconds without the dynamic distance potential field method.}
	\label{fig:Ecke2}
\end{figure}

Without dynamic potential after 90 seconds the situation looks like fig \ref{fig:Ecke2}

\begin{figure}[htbp]
  \center
	\includegraphics[width=0.30\textwidth]{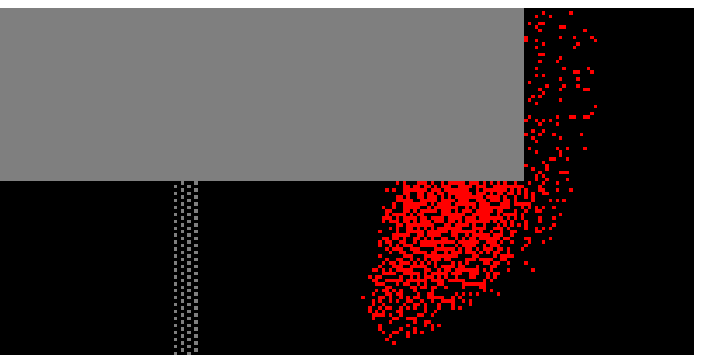}
	\includegraphics[width=0.30\textwidth]{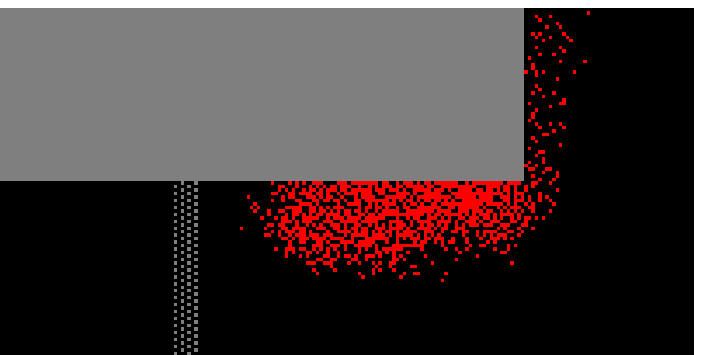}
	\includegraphics[width=0.30\textwidth]{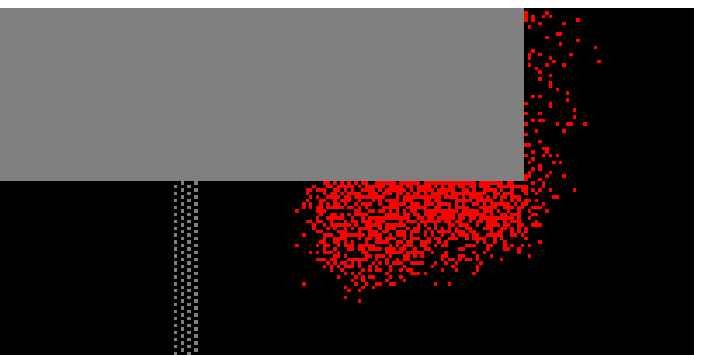}\\ \vspace{6pt}
	\includegraphics[width=0.30\textwidth]{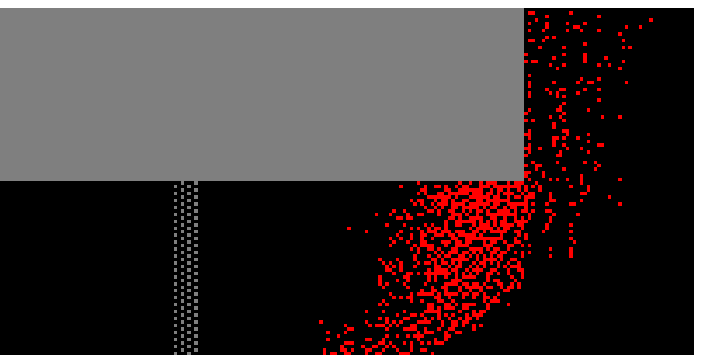}
	\includegraphics[width=0.30\textwidth]{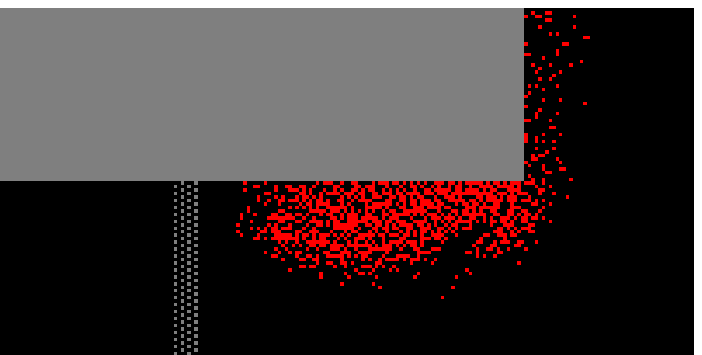}
	\includegraphics[width=0.30\textwidth]{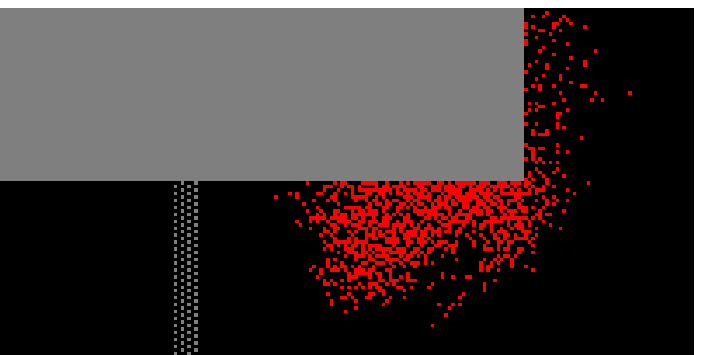}\\ \vspace{6pt}
	\includegraphics[width=0.30\textwidth]{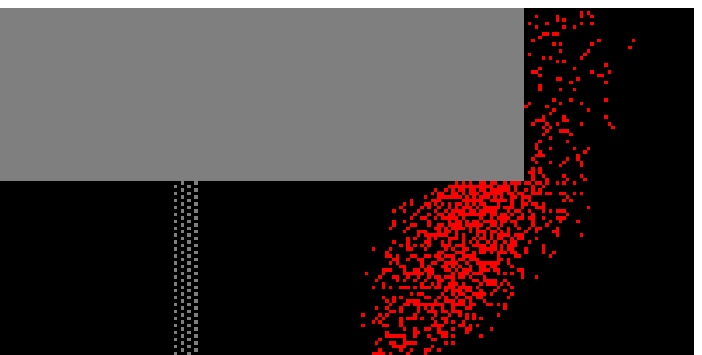}
	\includegraphics[width=0.30\textwidth]{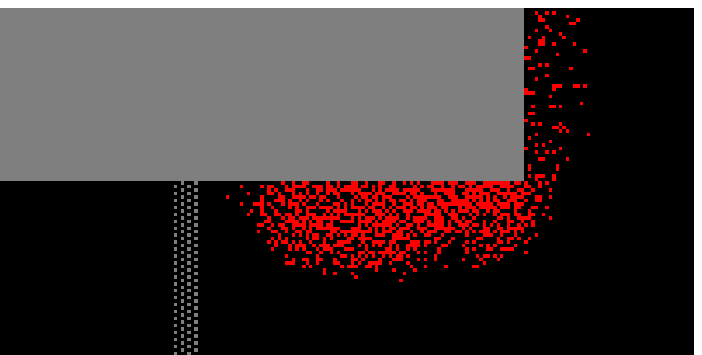}
	\includegraphics[width=0.30\textwidth]{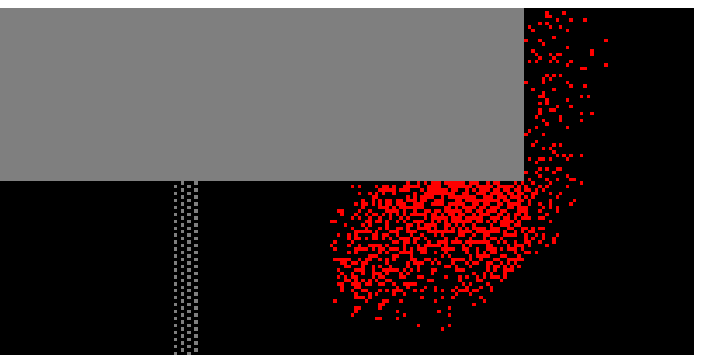}\\ \vspace{6pt}
	\includegraphics[width=0.30\textwidth]{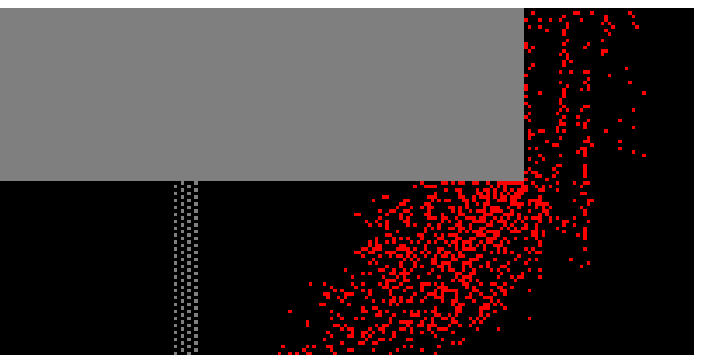}
	\includegraphics[width=0.30\textwidth]{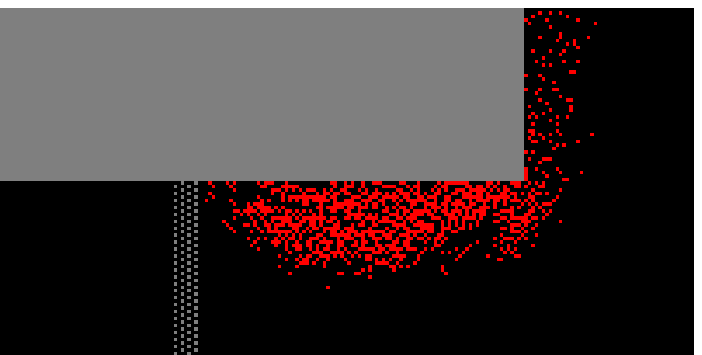}
	\includegraphics[width=0.30\textwidth]{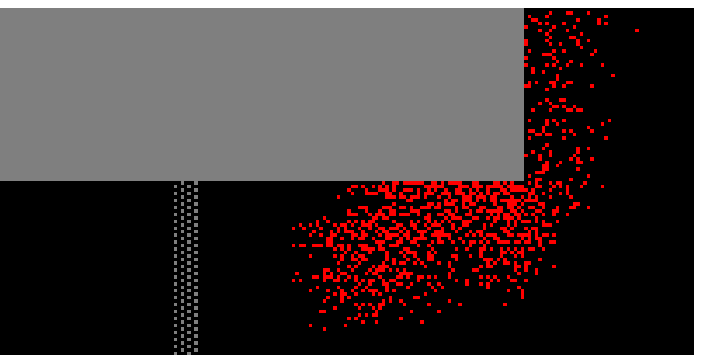}\\ \vspace{6pt}
	\includegraphics[width=0.30\textwidth]{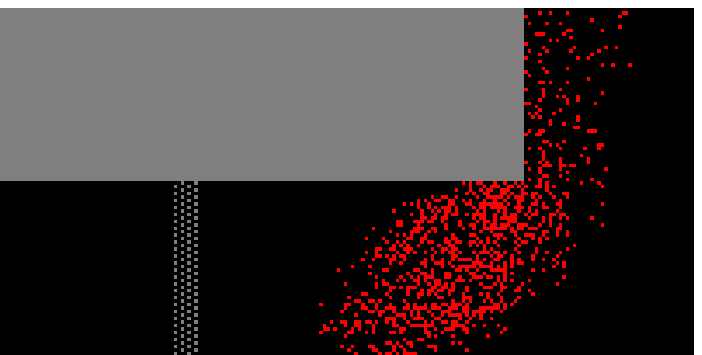}
	\includegraphics[width=0.30\textwidth]{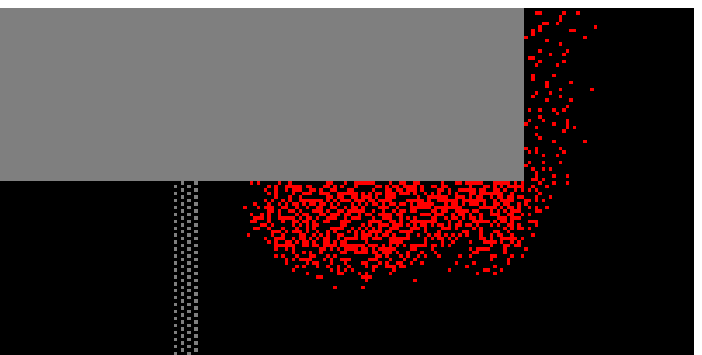}
	\includegraphics[width=0.30\textwidth]{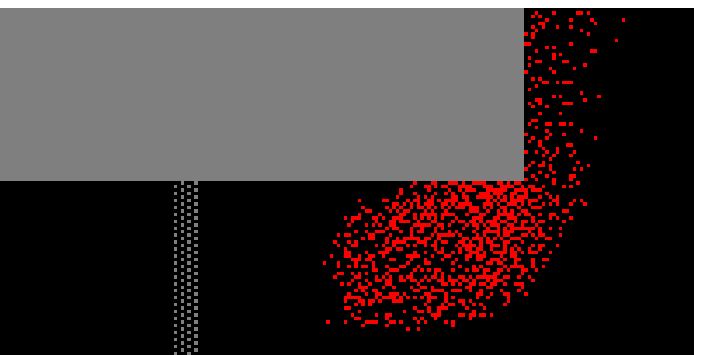}\\ \vspace{6pt}
	\includegraphics[width=0.30\textwidth]{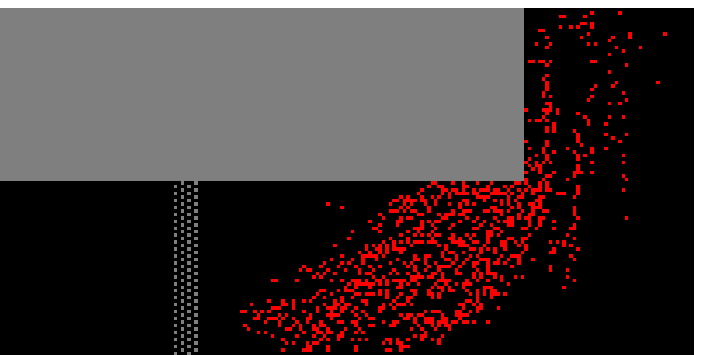}
	\includegraphics[width=0.30\textwidth]{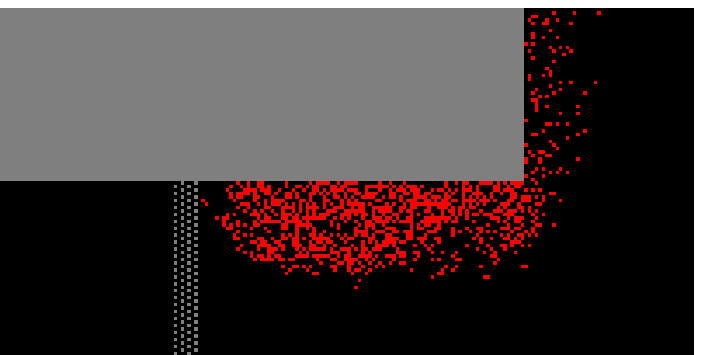}
	\includegraphics[width=0.30\textwidth]{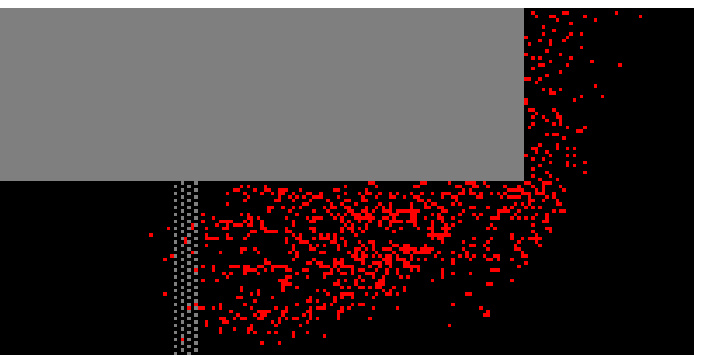}\\
	\caption{Situation after 90 seconds with the dynamic distance potential field method. The ordering of figures with regard to different methods and parameters is the same as in table \ref{tab:Ecke1}.}
	\label{fig:Ecke3}
\end{figure}
Figure \ref{fig:Ecke3} shows that the influence on the shape of the crowd is strongest if the Manhattan metric is used and weakest with the Chebyshev metric. This is not astonishing, as in this case the possibility to flood diagonally over just common corners and not only over common edges quickly can erase the effect of extra points in the potential due to agents. Furthermore one can see that large $k_{Sdyn}$ tend to exhibit artifacts in form of unrealistic density waves and this in the end leads to a re-increase of the evacuation time, when $k_{Sdyn}$ is too large.

To check the direction dependence of the method, the geometry was rotated by 45 degree. A direction dependency can come into play by the anisotropy of the speed neighborhood \cite{Kretz2006k} and the anisotropy of the dynamic distance potential field method. Without dynamic distance potential field the process took between 303 and 316 rounds, at an average of $309.0\pm 2.5$ rounds. With $k_{Sdyn}=1.0$ and $s_{add}=10$ it took with Manhattan metric $176.9\pm 1.8$, Chebyshev metric $235.2\pm 2.2$, and V1 metric $173.4\pm 1.7$. 

\subsection{A Crowd Moving Along a Straight Corridor} \label{sec:Straight}

As the dynamic distance potential field method can lead to an increase of the walking speed, one has to show that the reduction of the evacuation time as exhibited in table \ref{tab:Ecke1} is a consequence of the agents utilizing space at the corner more efficiently and not a consequence of them just walking faster. Therefore simulations have been done, where he agents did not have to move around the corner but walk on straight forward. Anything else was left as shown in figure \ref{fig:Ecke1}. Without dynamic distance potential field the process took between 113 and 118 rounds, at an average of 113.5 $\pm $ 1.6 rounds. The results for the various metrics and different parameters is shown in \ref{tab:Straight}.

\begin{table}[htbp]
\center
\small
\begin{tabular}[htbp]{c|ccc}
$k_{Sdyn}$ / $s_{add}$ &Manhattan       & Chebyshev                & V1\\ \hline
1.0 / 2     &  116.6 $\pm$ 1.4 (114, 121)& 113.8 $\pm$ 1.2 (112, 117)& 114.1 $\pm$ 1.5 (111, 118) \\
3.0 / 2     &  133.0 $\pm$ 1.6 (130, 139)& 114.8 $\pm$ 1.7 (112, 122)& 114.6 $\pm$ 1.3 (112, 118) \\
1.0 / 4     &  118.1 $\pm$ 1.5 (116, 122)& 114.1 $\pm$ 1.5 (112, 120)& 113.8 $\pm$ 1.2 (111, 117) \\
3.0 / 4     &  137.3 $\pm$ 1.5 (135, 144)& 115.3 $\pm$ 1.6 (113, 121)& 114.9 $\pm$ 1.4 (112, 118) \\
1.0 / 10    &  119.3 $\pm$ 1.5 (117, 125)& 114.4 $\pm$ 1.4 (112, 119)& 113.7 $\pm$ 1.3 (111, 117) \\
10.0 / 10   &  154.1 $\pm$ 1.9 (151, 163)& 115.8 $\pm$ 1.5 (113, 120)& 116.6 $\pm$ 1.5 (114, 123) \\
\end{tabular}
\caption{Number of rounds until the last agent has walked through the corridor and left the simulation}
\label{tab:Straight}
\normalsize
\end{table}

\begin{figure}[htbp]
  \center
	\includegraphics[width=0.41\textwidth]{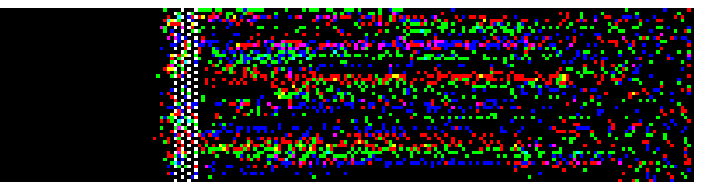}
	\caption{The situation in three subsequent rounds: 89 (red), 90 (green), and 91 (blue) when Manhattan metric is used. It can be seen that there are specific lanes in each round and that the lanes of immediately consequent rounds are almost entirely separated (there is no yellow or cyan), but that there is one lane in round 91 at a position where there was one in round 89 (magenta). }
	\label{fig:Straight1}
\end{figure}

The results from \ref{tab:Straight} show two remarkable things: first off, the dynamic distance potential field does not speed up the process. Second, the combination of Manhattan metric with large $k_{Sdyn}$ can -- at least in this case with motion along the axis of discretization -- lead to considerable delays. The reason is not that each single agent is slowed down directly, but that small density variations build up as the agents simultaneously move to lanes where in front of them relatively few other agents are located. This position is evaluated as being particularly bad and the agents move to the side in subsequent rounds. But it's not the sidewards motion that leads to delays but the locally high density right in front of them toward the destination. Figure \ref{fig:Straight1} illustrates this effect. As this under any circumstances can only be an unwanted effect the use of the Manhattan metric demands greater care in setting $k_{Sdyn}$ and after the simulation is carried out a check, if or if not this artificial lane formation appeared at some time. It is therefore better to not use the Manhattan metric at all or to replace $k_{Sdyn}$ by another parameter $a$ from which $k_{Sdyn}$ calculates for example as
\begin{equation}
k_{Sdyn}=\frac{k_S}{1+a^2}.
\end{equation}
In this case $k_{Sdyn}$ can only be smaller than $k_S$. 

The Chebyshev metric on the contrary only has a minor influence on both the shape of a crowd moving around a corner and the evacuation time. What remains is the method of Variant 1. Nevertheless for the rest of the contribution consider all three metrics will be considered to check and consolidate the impression that there are good reasons to use variant 1 (V1).

\subsection{A Crowd Doing a U-Turn Around a Wall}
Next the agents from the same starting positions had to do a u-turn around a 40 cm wide wall. Compared to figure \ref{fig:Ecke1} the unaccessible (white) area above the starting corridor were accessible and the destination was on the other side of the wall ``above" the flow restricting columns. Without dynamic distance potential field the process took between 461 and 476 rounds, at an average of 468.75 $\pm $ 3.1 rounds. The results for the various metrics and different parameters are shown in \ref{tab:U-Turn}.

\begin{table}[htbp]
\center
\small
\begin{tabular}[htbp]{c|ccc}
$k_{Sdyn}$ / $s_{add}$ &Manhattan        & Chebyshev                 & V1\\ \hline
1.0 / 2      & 335.7 $\pm$ 2.1 (330, 341)& 385.9 $\pm$ 2.1 (382, 392)& 375.3 $\pm$ 2.7 (370, 384) \\ 
3.0 / 2      & 273.5 $\pm$ 2.1 (268, 279)& 331.7 $\pm$ 2.2 (325, 338)& 313.2 $\pm$ 2.2 (309, 322) \\ 
1.0 / 4      & 260.5 $\pm$ 1.8 (255, 265)& 334.9 $\pm$ 2.3 (329, 342)& 299.5 $\pm$ 2.1 (294, 304) \\ 
3.0 / 4      & 234.9 $\pm$ 2.0 (231, 240)& 302.1 $\pm$ 2.1 (297, 308)& 249.1 $\pm$ 1.8 (245, 255) \\ 
1.0 / 10     & 225.0 $\pm$ 2.1 (219, 231)& 315.4 $\pm$ 2.4 (310, 320)& 249.8 $\pm$ 1.8 (246, 256) \\ 
10.0 / 10    & 233.0 $\pm$ 2.4 (229, 239)& 298.5 $\pm$ 3.2 (291, 307)& 235.1 $\pm$ 2.4 (228, 243) \\ 
\end{tabular}
\caption{Number of rounds until the last agent has done the u-turn and left the simulation}
\label{tab:U-Turn}
\normalsize
\end{table}

These results reveal a reduction of the evacuation time by 50\% for some of the parameters compared to the case without dynamic distance potential field. Again the Chebyshev metric has the smallest and the Manhattan metric the largest effect. The strength of artificial effects strongly depends on $k_{Sdyn}$ and almost not on $s_{add}$.

\subsection{The Fundamental Diagram}
\begin{figure}[htbp]
  \center
	\includegraphics[width=0.90\textwidth]{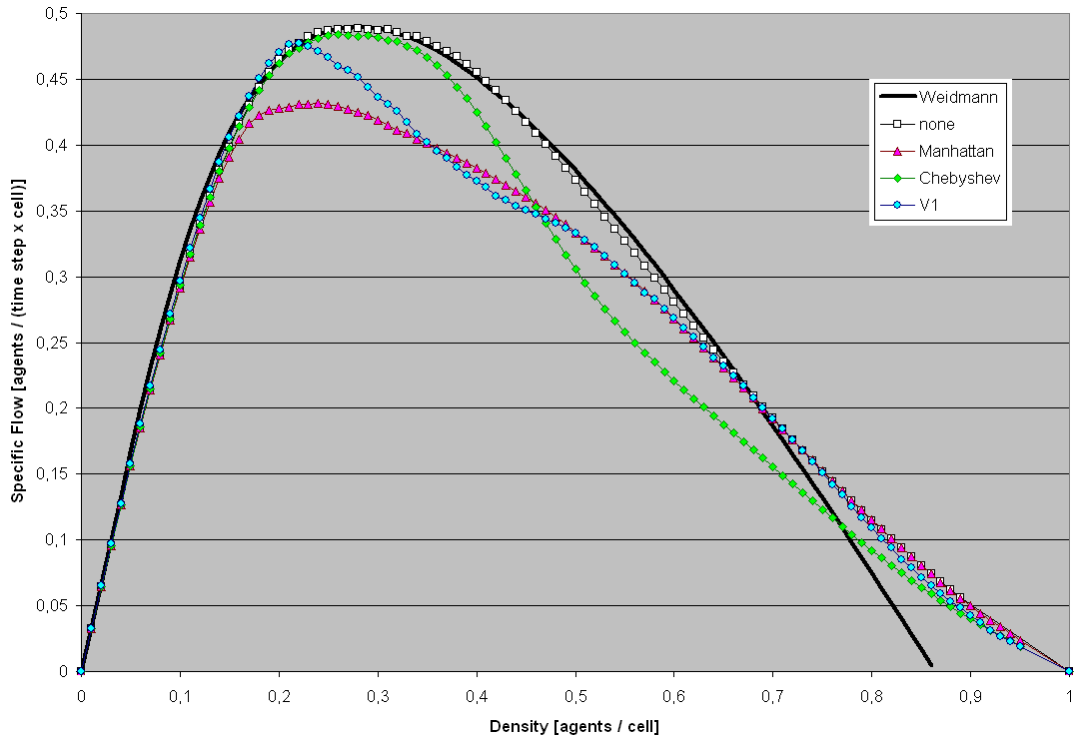}
	\caption{Theoretical fundamental diagram (fd) by Weidmann (see equation (\ref{eq:Weidmann})), fd from the F.A.S.T. model without dynamic distance potential field, and with the dynamic distance potential fields using Manhattan, Chebyshev, and V1 metric. All F.A.S.T. simulations have been carried out with $k_S=1.2$, those with dynamic distance potential field with $k_{Sdyn}=1.0$ and $s_{add}=10$. All other $k$ were set to zero. As one round equals one second and one cell has a size of 0.4 X 0.4 sqm, the density is shown in fractions of 6.25 agents per sqm, and the specific flux in fractions of 2.5 agents per meter and second.}
	\label{fig:FD}
\end{figure}

In \cite{Kretz2007a} it was shown that the F.A.S.T. model can reproduce Weidmann's fundamental diagram \cite{Weidmann1993} quite well with the rather simple set of parameters $v_{max}=4$, $k_S=1.2$, and $k_{other}=0.0$. Therefore the fundamental diagram was calculated with these parameters and the dynamic distance potential field added. And it was calculated using the same geometry: a 4 meters (10 cells) wide ring shaped corridor with an outer radius of 500 cells. There were four destinations at 0, 90, 180, and 270 degree making the agents walk around. The reason for the selection of this geometry once was that it would show remaining anisotropies and that in principle one could implement this in an experiment, which is not possible in a straight corridor with periodic boundary conditions. The results are shown in figure \ref{fig:FD}.

There are two possible reasons for the reduction of the flow at the intermediate densities: a) the deviation of the dynamic metric from the Euclidean metric and b) effects of percolation \cite{Stauffer1994}. To decide about this an additional fundamental diagram was calculated with four additional intermediate distances at 45, 135, 225, and 315 degree and V1 metric. As V1 produces errors only at corners, with these additional intermediate destinations, the average error of the dynamic distance potential field was reduced. In effect the flow actually increased, but the maximum effect was an increase by 0.02 agents per round and cell at a density of 0.3 agents per cell. So, by far the largest share of flow reduction must be due to the percolation of unoccupied cells vanishing with increasing density. This is supported by the fact that both for Manhattan and Chebyshev metric the largest deviations from the case without dynamic distance potential field appear at densities about 0.06 agents per cell smaller than the percolation threshold of agents (on an infinite lattice) with regard to the metrics' neighborhood. Analytical investigations of this phenomenon are difficult not only due to the geometry, but as one cannot assume in advance that the agents' positions are uncorrelated.

Annotation: Weidmann's theoretical fundamental diagram follows the formula
\begin{equation}
j = \rho \cdot v_{free} \left(1-\exp\left(-\gamma\left(\frac{1}{\rho}-\frac{1}{\rho_{max}}\right)\right)\right)
\label{eq:Weidmann}
\end{equation}
with flow $j$, density $\rho$, free speed $v_{free}=1.34$ m/s, maximum density $\rho_{max}=5.4$ 1/sqm, and the gauge constant $\gamma=1.913$ 1/sqm.

\subsection{A Part of a Stadium} \label{sec:Stadium}
\begin{figure}[htbp]
  \center
	\includegraphics[width=306.5pt]{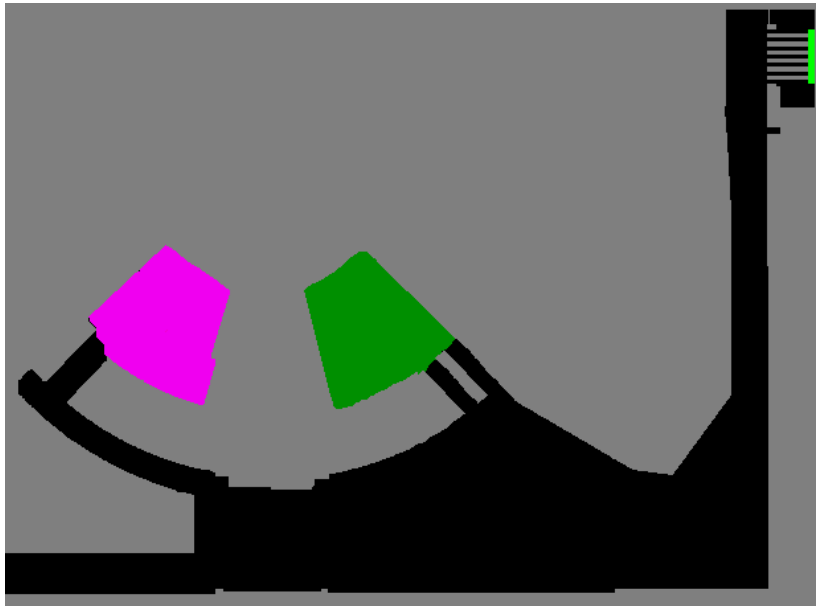}
	\caption{Geometry and agents' starting positions for the simplified stadium test simulation. On the magenta and dark green areas two times 5,000 agents started their egress and had to walk to the light green area on the upper right.}
	\label{fig:Stadium1}
\end{figure}

To go beyond the elementary geometries investigated so far a scenario with 10.000 agents moving through a simplified stadium-like geometry was simulated. Figure \ref{fig:Stadium1} shows the geometry and initial positions of the agents. The average time of 100 runs until the last agent had left the scenario without the dynamic distance potential field method was $2956.8 \pm 8.1$ rounds (seconds). The average of individual egress times was $1579.4\pm 3.5$ rounds. The results for the process using the various metrics and different parameters are shown in tables \ref{tab:Stadium1} and \ref{tab:Stadium2}.

\begin{table}[htbp]
\center
\normalsize
\begin{tabular}[htbp]{c|ccc}
$k_{Sdyn}$ / $s_{add}$ & Manhattan & Chebyshev & V1     \\ \hline
1.0 / 2                & 2096.2    & 2228.2    & 2040.4 \\
3.0 / 2                & 1785.9    & 1900.1    & 1697.4 \\
10.0 / 2               & 1739.6    & 1854.3    & 1753.8 \\
1.0 / 4                & 1673.9    & 1905.9    & 1620.6 \\
3.0 / 4                & 1568.1    & 1823.8    & 1424.4 \\
1.0 / 10               & 1503.5    & 1861.5    & 1417.5 \\
10.0 / 10              & 1646.5    & 1959.0    & 1962.7 \\
\end{tabular}
\caption{Average of 100 number of rounds until the last agent has left the simplified stadium simulation.}
\label{tab:Stadium1}
\normalsize
\end{table}

\begin{table}[htbp]
\center
\normalsize
\begin{tabular}[htbp]{c|ccc}
$k_{Sdyn}$ / $s_{add}$ &Manhattan & Chebyshev & V1     \\ \hline
1.0 / 2                & 1147.6   & 1238.0    & 1138.4 \\
3.0 / 2                & 1003.2   & 1084.8    &  980.5 \\
10.0 / 2               &  992.0   & 1067.4    & 1045.0 \\
1.0 / 4                &  950.6   & 1094.8    &  936.4 \\
3.0 / 4                &  916.6   & 1055.2    &  856.2 \\
1.0 / 10               &  886.2   & 1077.6    &  842.8 \\
10.0 / 10              &  957.3   & 1109.1    & 1119.7 \\
\end{tabular}
\caption{Average of 100 individual egress times in the simplified stadium simulation.}
\label{tab:Stadium2}
\normalsize
\end{table}

\begin{figure}[htbp]
  \center
	\includegraphics[width=306.5pt]{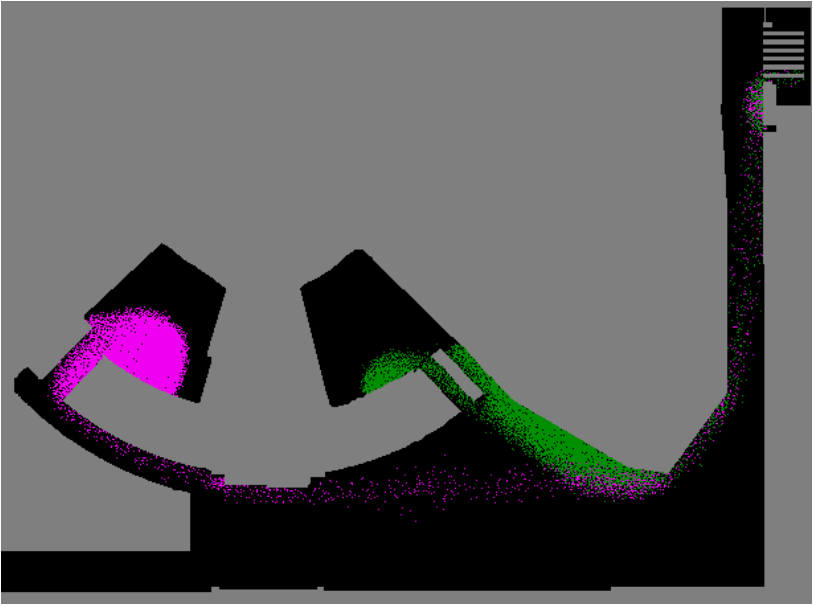}
	\caption{The situation after 400 rounds without dynamic distance potential.}
	\label{fig:Stadium2}
\end{figure}

\begin{figure}[htbp]
  \center
	\includegraphics[width=306.5pt]{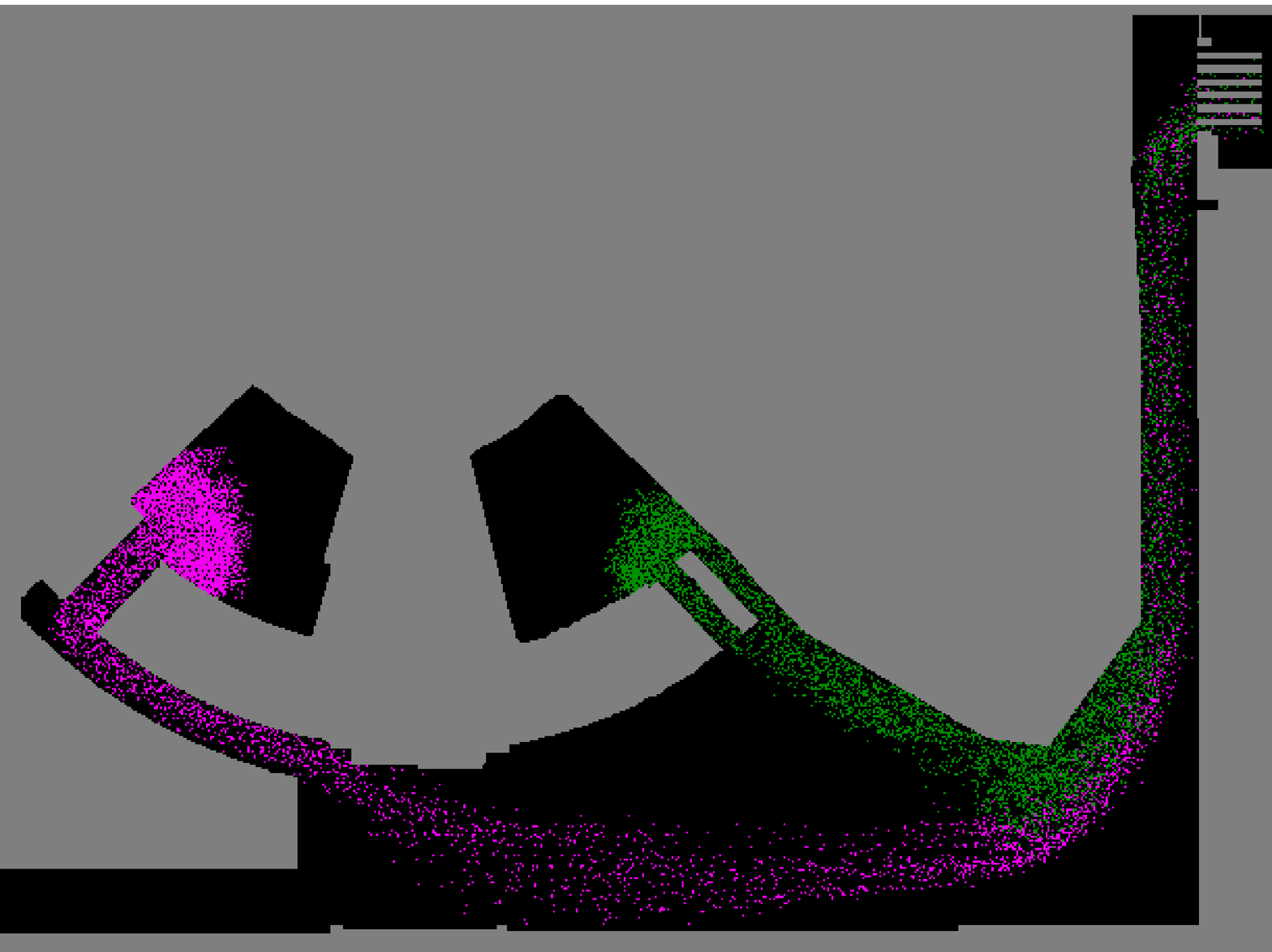}
	\caption{The situation after 400 rounds with Manhattan metric.}
	\label{fig:Stadium3}
\end{figure}

\begin{figure}[htbp]
  \center
	\includegraphics[width=306.5pt]{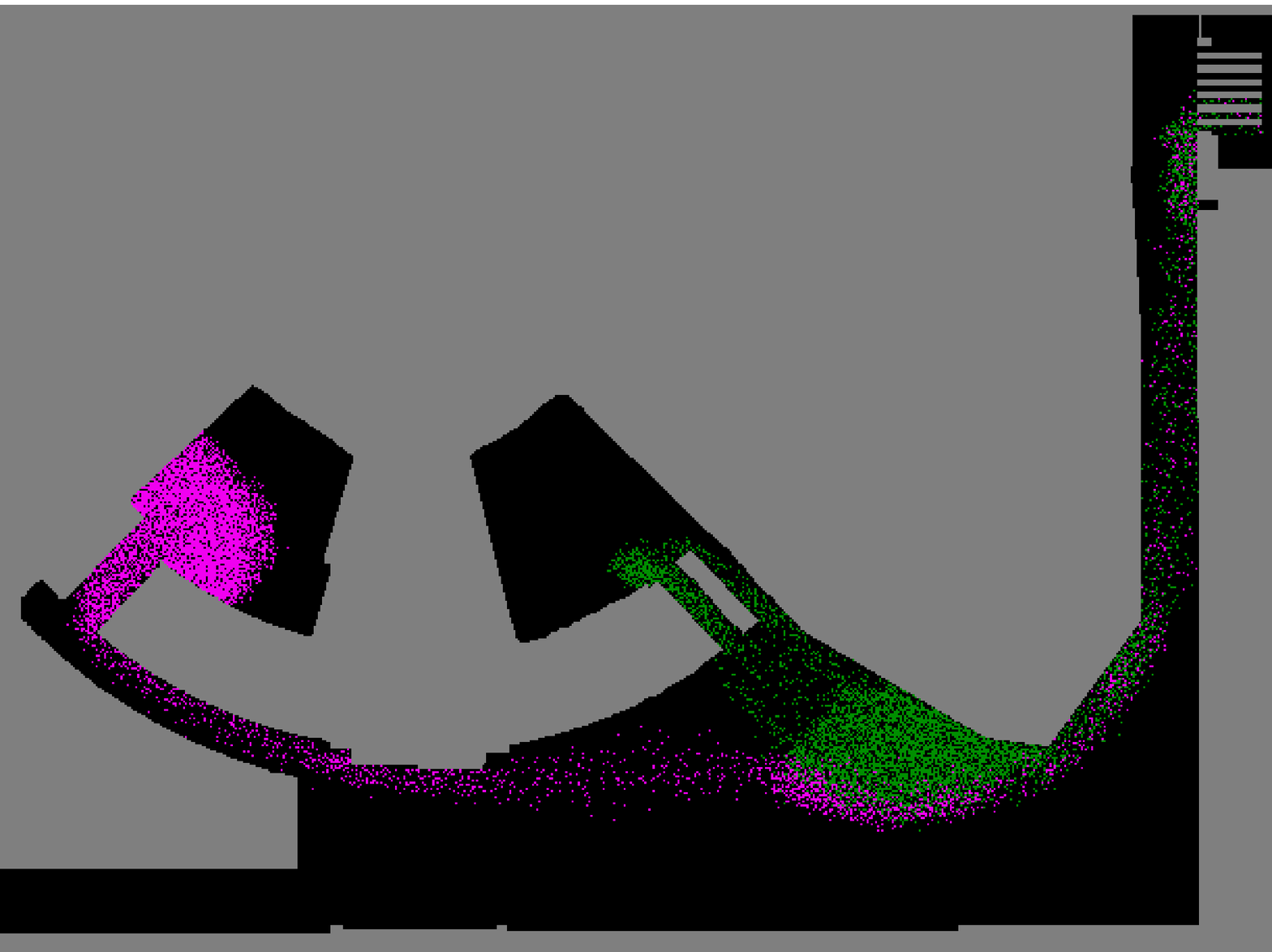}
	\caption{The situation after 400 rounds with Chebyshev metric.}
	\label{fig:Stadium4}
\end{figure}

\begin{figure}[htbp]
  \center
	\includegraphics[width=306.5pt]{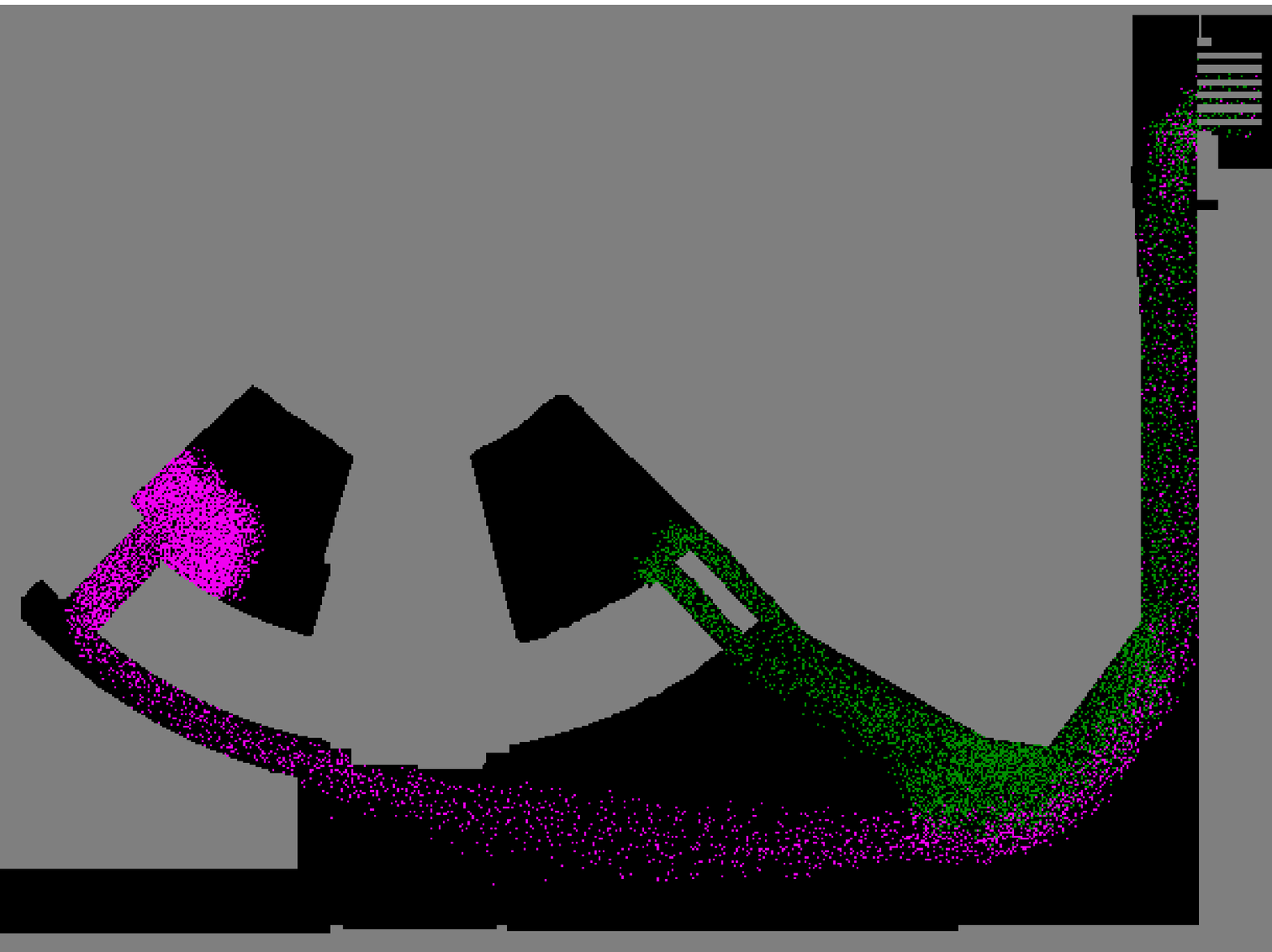}
	\caption{The situation after 400 rounds with V1 metric.}
	\label{fig:Stadium5}
\end{figure}

\begin{figure}[htbp]
  \center
	\includegraphics[width=306.5pt]{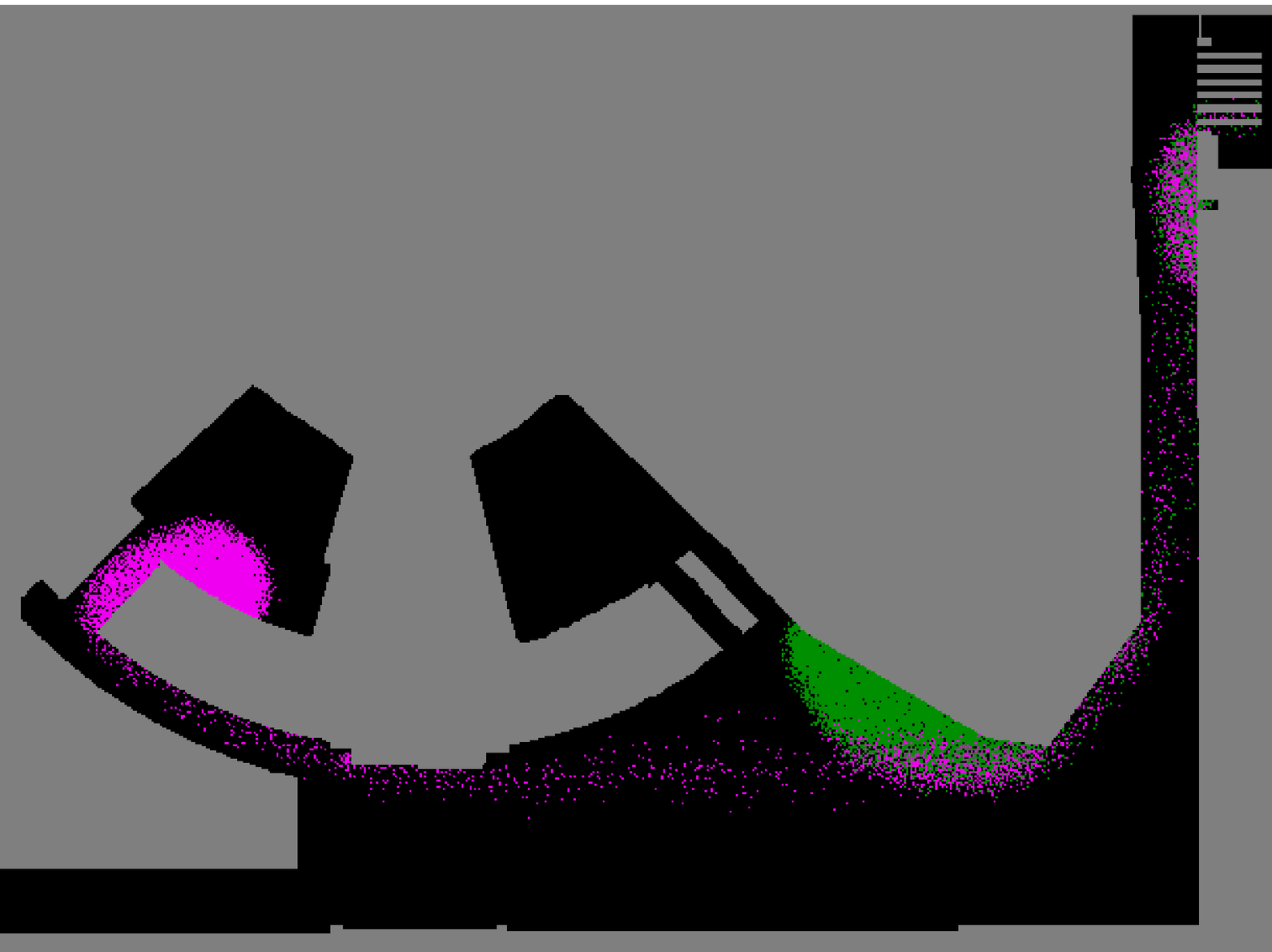}
	\caption{The situation after 800 rounds without dynamic distance potential.}
	\label{fig:Stadium6}
\end{figure}

\begin{figure}[htbp]
  \center
	\includegraphics[width=306.5pt]{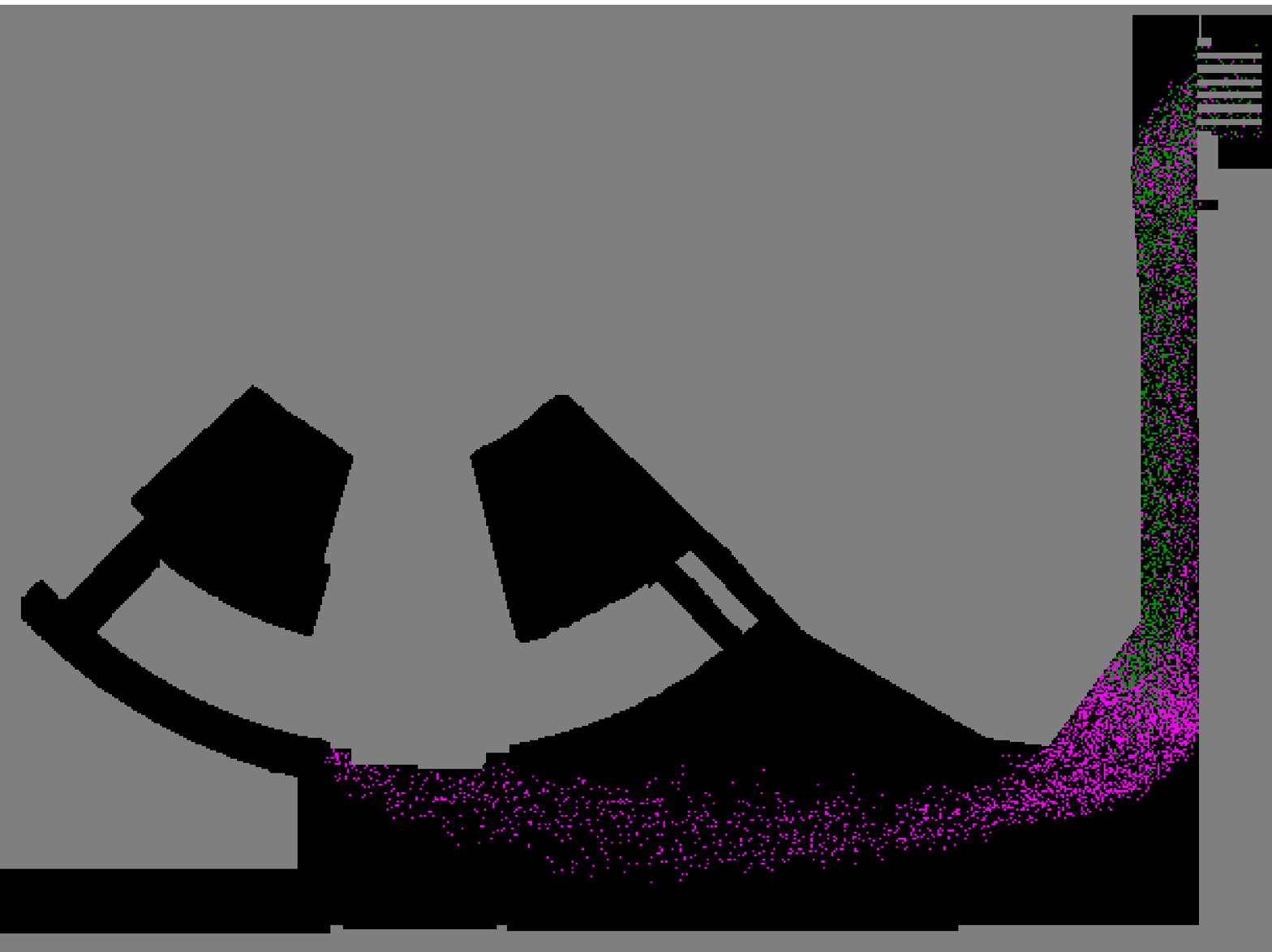}
	\caption{The situation after 800 rounds with Manhattan metric.}
	\label{fig:Stadium7}
\end{figure}

\begin{figure}[htbp]
  \center
	\includegraphics[width=306.5pt]{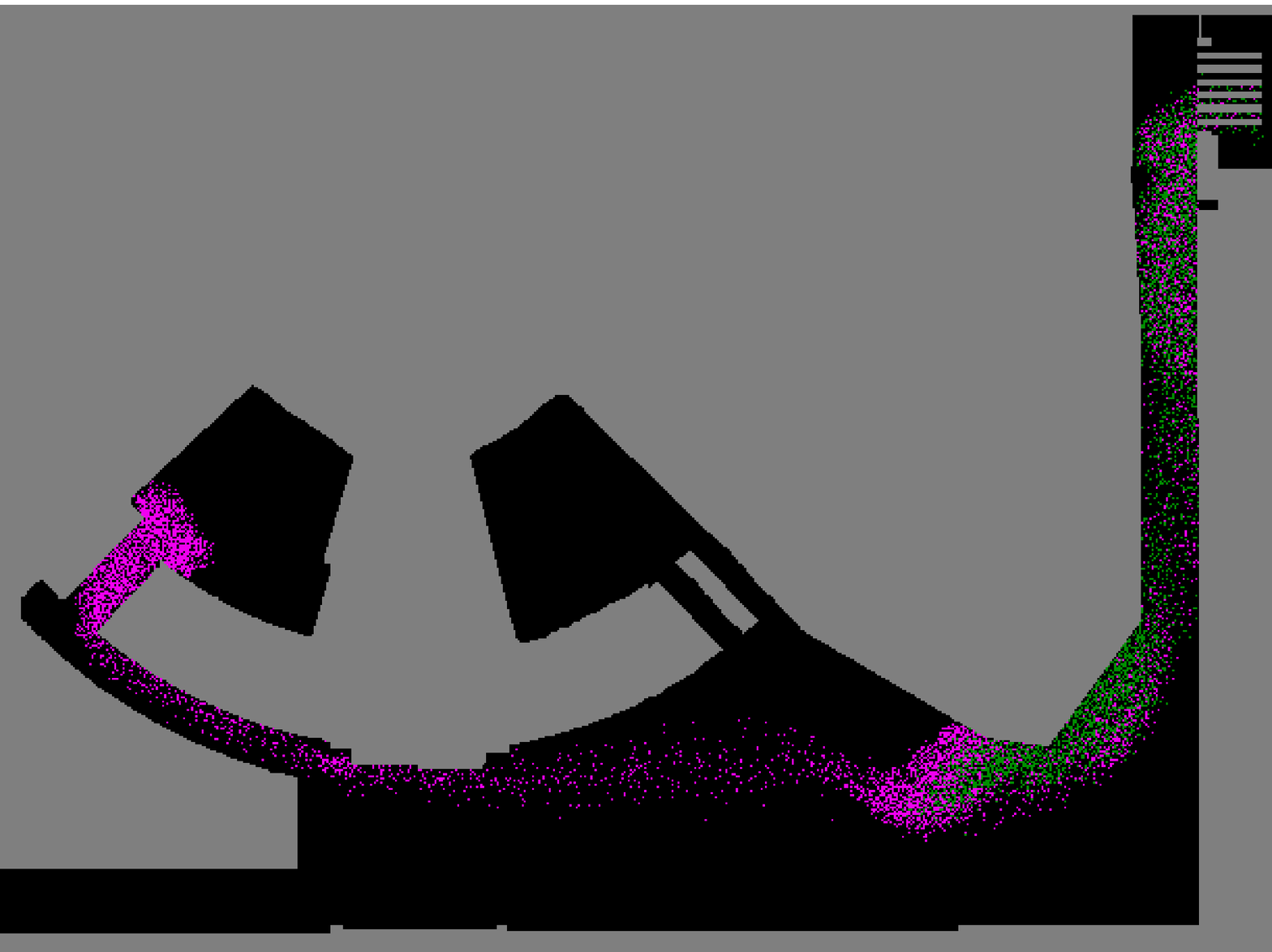}
	\caption{The situation after 800 rounds with Chebyshev metric.}
	\label{fig:Stadium8}
\end{figure}

\begin{figure}[htbp]
  \center
	\includegraphics[width=306.5pt]{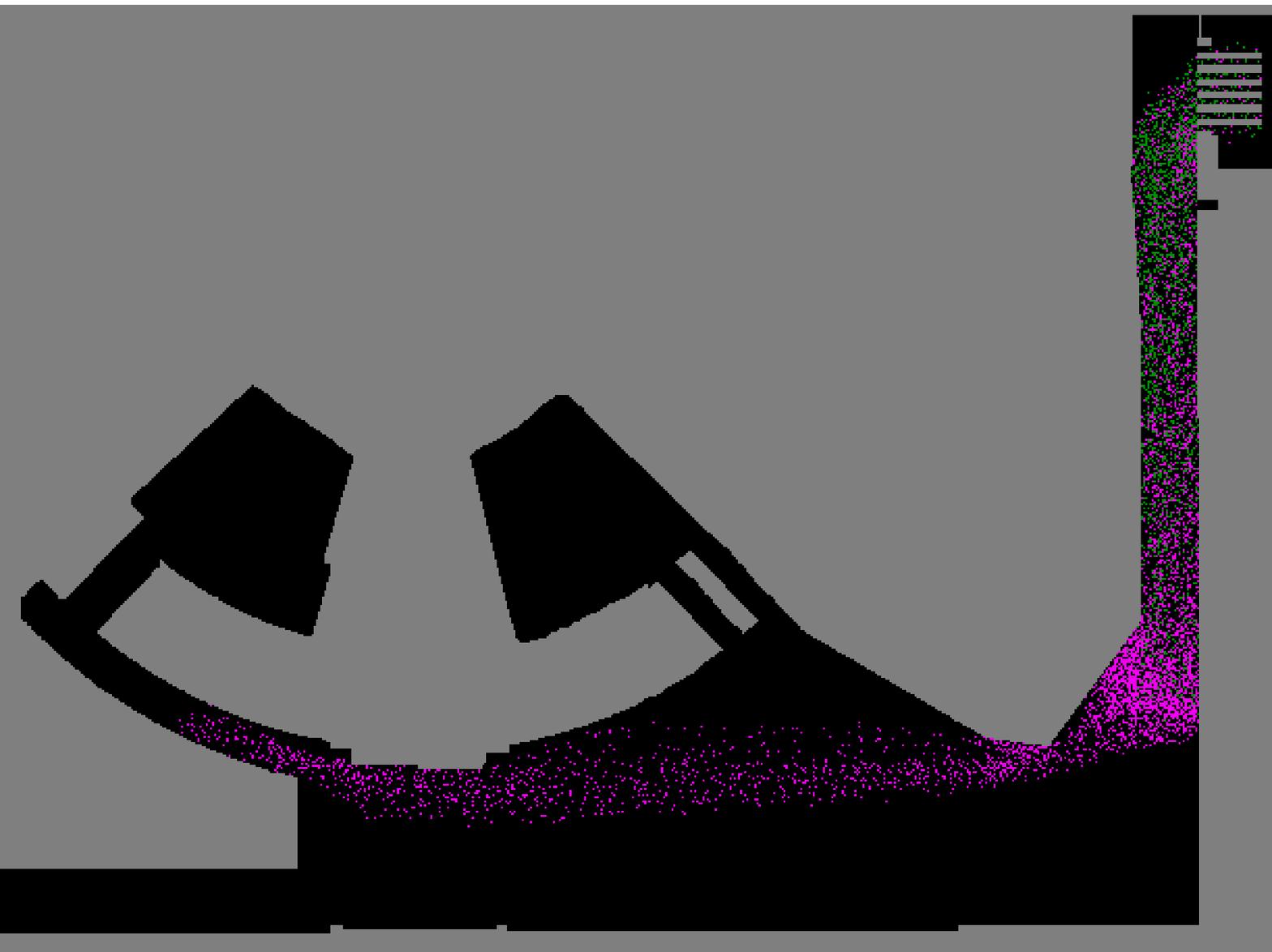}
	\caption{The situation after 800 rounds with V1 metric.}
	\label{fig:Stadium9}
\end{figure}

\begin{figure}[htbp]
  \center
	\includegraphics[width=306.5pt]{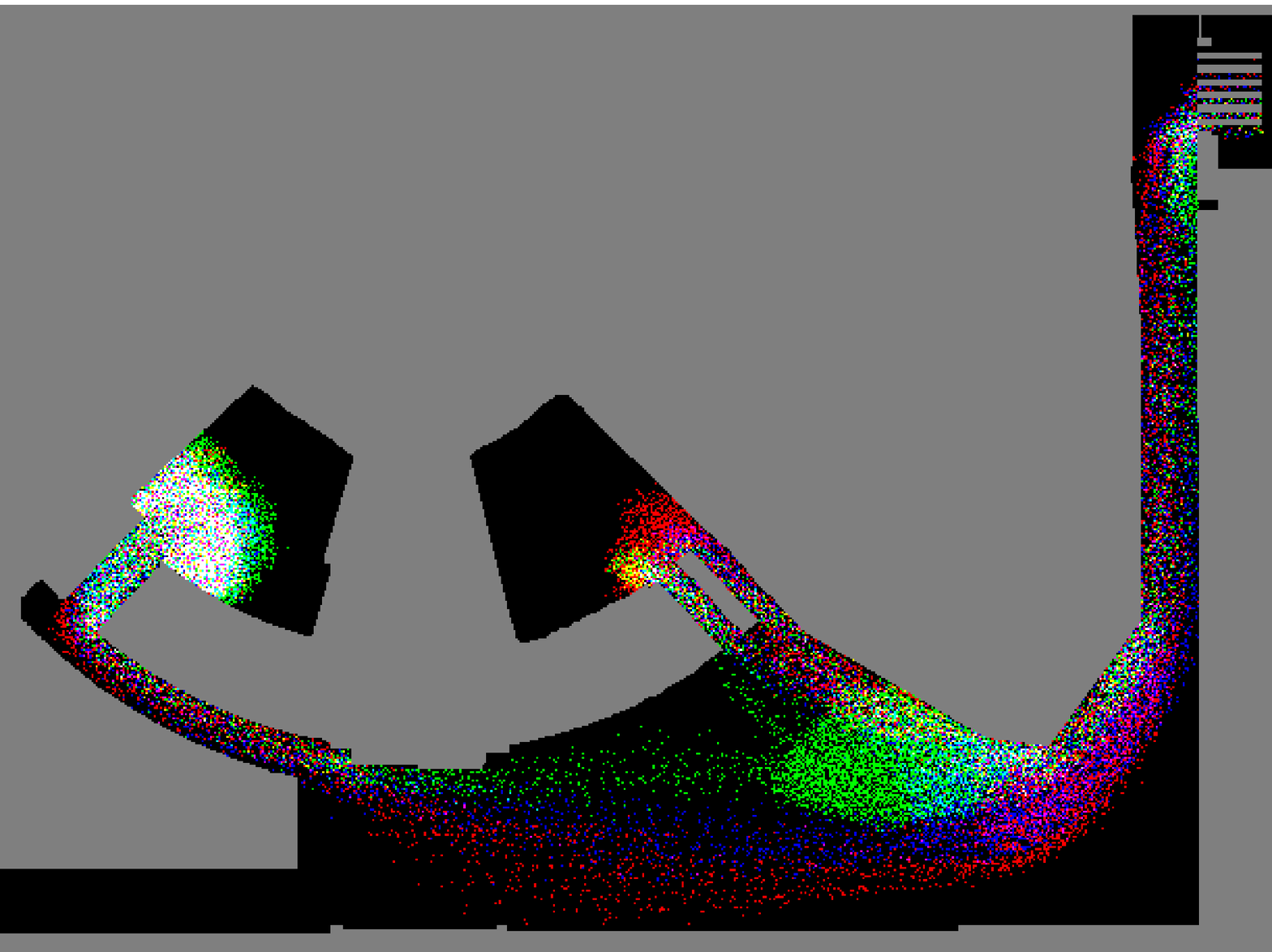}
	\caption{The situation after 400 rounds with Manhattan (red), Chebyshev (green), and V1 (blue) metric.}
	\label{fig:Stadium10}
\end{figure}

\begin{figure}[htbp]
  \center
	\includegraphics[width=306.5pt]{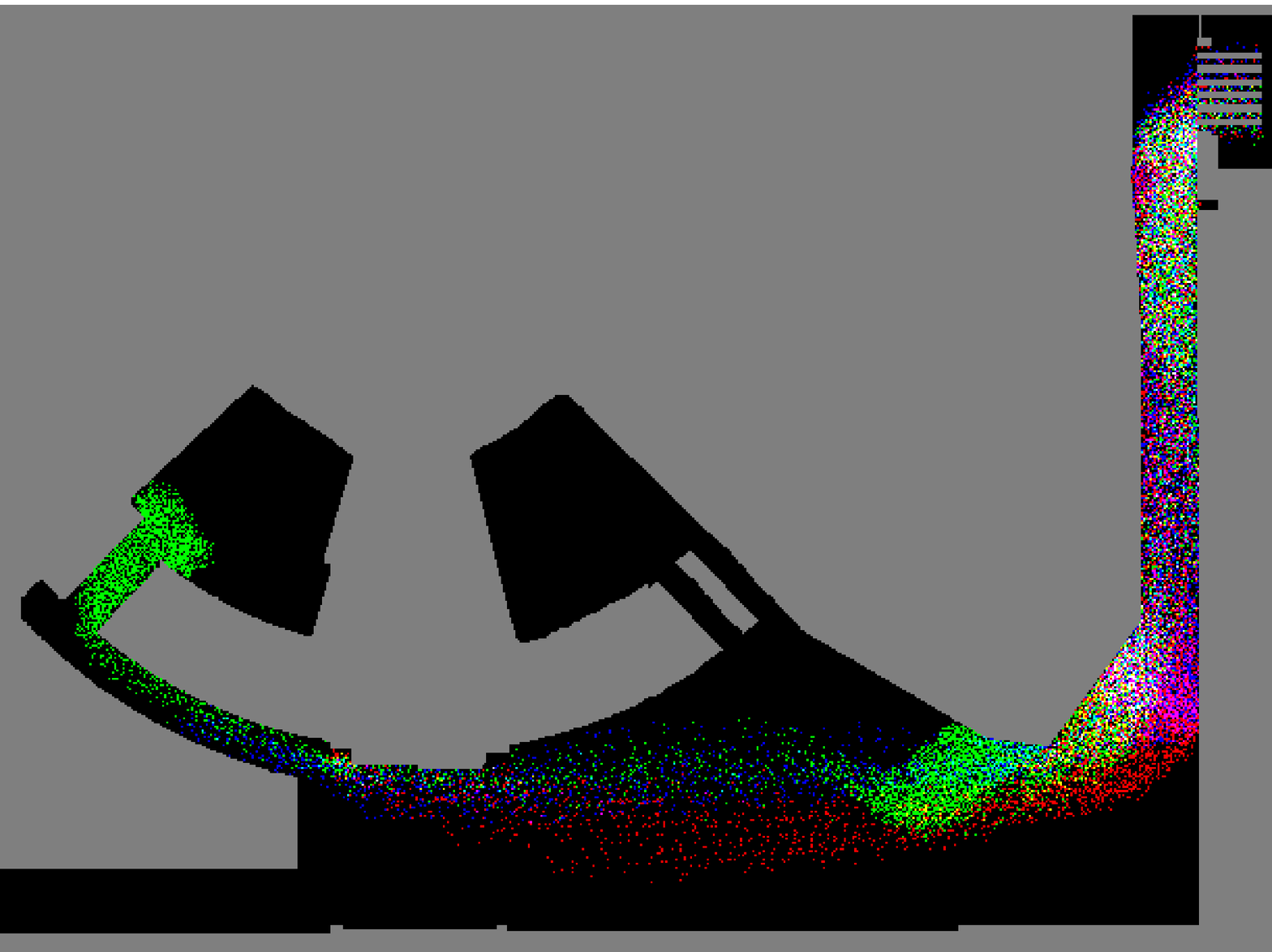}
	\caption{The situation after 800 rounds with Manhattan (red), Chebyshev (green), and V1 (blue) metric.}
	\label{fig:Stadium11}
\end{figure}

The first fact to realize when looking at tables \ref{tab:Stadium1} and \ref{tab:Stadium2} is that method V1 leads to the smallest total and individual evacuation times for $k_{Sdyn}=1.0$ and $k_{Sdyn}=3.0$. With V1 the evacuation times are larger for $k_{Sdyn}=10.0$ than for the next smaller $k_{Sdyn}$ with identical $s_{add}$. In total the effects show most pronounced -- and also "best" -- for $s_{add}=10$ and unrealistic artifacts are smallest for $k_{Sdyn}=1.0$. Therefore in figures \ref{fig:Stadium2} to \ref{fig:Stadium11} the case without dynamic distance potential field is compared at $t=400$ and $t=800$ to the three variants using this choice of parameters. It's interesting to take a look at
\begin{itemize}
\item The shape of the crowd waiting on the initial area.
\item The vertical position of the stream walking from left to right.
\item The jam at the corner, where the main moving direction changes.
\item The distribution of agents in the final corridor.
\item The utilization of the gates at the end.
\end{itemize}

For all of these issues things are either looking best using V1 or there is no difference between the three metric variants. In any case it's better to make use of a dynamic distance potential field.

Apart from the simulations for which the results are shown in tables \ref{tab:Stadium1} and \ref{tab:Stadium2} the following variants have been simulated:
\begin{itemize}
\item All three metrics with $k_{Sdyn}=1.0$ and $s_{add}=100$. The results did not in any aspect (evacuation time, individual egress time, shape of crowd, etc.) deviate significantly or recognizably from the ones with $s_{add}=10$. This confirms that with increasing $s_{add}$ there is a saturation of the influence of this parameter.
\item V1 again with $k_{Sdyn}=1.0$ and $s_{add}=100$, but now the field values for $S_{dyn}$ and $S_{dyn}^t$ were not rounded to integer values after taking the square root in the calculation process. This just as well did not create recognizable macroscopic effects.
\item V1 again with $k_{Sdyn}=1.0$ and $s_{add}=100$, but now the $\Delta S_{dyn}$ values for the current position of a agent were ignored (artificially set to zero). This created a very unrealistic behavior, as as long as an agent had other agents in some not too large distance ahead of him he preferred his current position a lot over any other and remained standing still.
\item Without dynamic distance potential field, but with $k_W>0$ and $k_I>0$, so the static repulsion from walls and inertia were switched on. With that the evacuation time could be reduced. With $k_W>0$ but $k_I=0$ there was only a minor effect of a reduction of 30 rounds at $k_W=0.5$. From all parameter combinations investigated the largest effect was seen with $k_W=0$ and $k_I=2.0$ as the average evacuation time reduced to 2026.5 rounds. However, this cannot be attributed to the jamming crowd at the corner making following agents detour, but to a wider spread of the stream of agents coming from the upper source due to inertia. This spread would have taken place just as well, if there had not been a jam at all at the corner. In this case the increased inertia would have led to an increase of evacuation time. Thus, the behavior induced by the parameters was independent of the situation and therefore only accidentally led to a reduction of evacuation time.
\end{itemize}

\subsection{Computation Times}
The stadium was simulated for all three metrics with $s_{add}=1$, taking the effect from the method. The computation time is compared to the average simulation time of 6:00 with the calculation for the method switched off. The results are shown in table \ref{tab:CT}
\begin{table}[htbp]
\center
\normalsize
\begin{tabular}[htbp]{c|ccc}
                         & Manhattan & Chebyshev & V1\\ \hline
with dyn. potential      &  6:30     & 6:48      & 7:18\\
relative increase				 &  8.3\%    & 13.3\%    & 21.6\% \\
\end{tabular}
\caption{Average computation time ([min:sec]) with the dynamic potential field method with $s_{add}=1$ and the relative increase compared to simulations without dynamic distance potential field method.}
\label{tab:CT}
\normalsize
\end{table}

\begin{figure}[htbp]
  \center
	\includegraphics[width=0.9\textwidth]{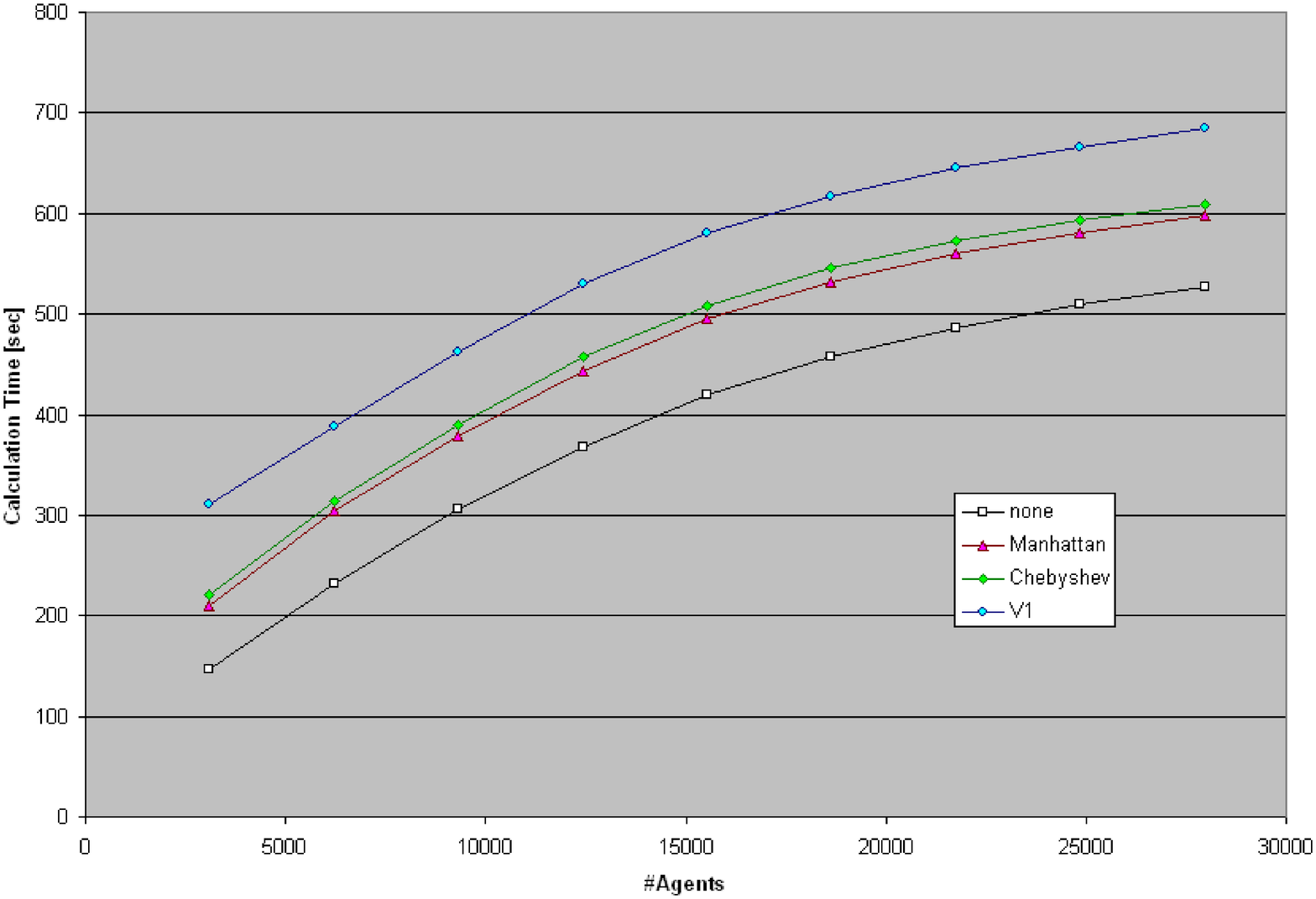}
	\caption{Average computation times (in seconds) for various densities in the fundamental diagram. 1100 rounds were simulated.}
	\label{fig:CalcTimes}
\end{figure}

When looking at the computation times during the calculation of the fundamental diagram, one can see the difference in computation time at different densities, respectively different numbers of agents in identical geometries. The computation times in figure \ref{fig:CalcTimes} were obtained with $s_{add}=1$, so the behavior of the agents did not change compared to the case without dynamic distance potential field. It can be seen that the absolute difference is about constant, while the relative one decreases with increasing number of agents.


The relative increase in computation time depends on the scenario, as the time for the calculation of the dynamic distance potential field only depends on the size of the area, but almost not on the number of agents in the simulation. This is a very positive fact compared to a na{\"i}ve ansatz of each agent having to observe any other to meet the goals as presented here. With such a na{\"i}ve ansatz the computation time would depend on the square of the number of agents. With the dynamic distance potential field the relative increase of computation time decreases with increasing global density, as typically the computation time depends at least linearly on the number of agents. As the method is most valuable to make agents evade large aggregations of other agents (jams) early, a user can switch off the method in simulations, where the local density is small everywhere and all the time. So, in the case when the method would imply the largest relative computation time increase, it is of no use anyway and can be skipped in the simulation algorithm. 

With regard to usage in other models, the relative increase in computation time is small, if the model needs much computation time. The computation time of the dynamic distance potential field is model independent. If a model has many time steps to simulate one second it might be possible to recalculate the dynamic distance potential field only once a simulation second, as within a second the situation never changes dramatically in agent dynamics. So, if there is a model with the same computational effort per time step as the F.A.S.T. model, but which is doing 20 time steps per simulation second, the increase in computation time might only be 1\%.

\section{Conclusions and Outlook}
The method proposed in this work proofed to do what was intended: the movement of the agents looked more natural in many cases. The agents showed more jam avoiding behavior, accepting longer paths around corners in exchange for higher walking speeds and thus the group of agents increased the total flow. The effect was strongest for the Manhattan metric. The smallest effect was achieved using the Chebyshev metric. Partly this may be due to the alignment of the geometry of most examples of this article to the axis of discretization. Without empirical check, it seems that the shape of the crowd is most natural using the V1 metric. As it is a combination of the two other methods effects on $S_{dyn}$ can reach far, but decrease with distance. It seems that setting $s_{add}$ to large values close to saturation (such that further increase does not change anything anymore) pronounces the positive effects most without bringing in negative drawbacks. It is possible, however, to choose too large values for $k_{Sdyn}$. Then the crowds will form unnatural shapes. In situations where the introduction of the method would not have been necessary the method only introduces small changes - at least if $k_{Sdyn}$ is reasonably small.

A metric investigated in \cite{Kretz2008c}, but not considered here, is V2, where the Moore neighborhood is used but a $\sqrt{2}$ is added for in the flood fill over corners. As a real-valued V1 metric seems to bring no changes, the integer-valuedness of the metric seems to be not necessary and V2 can also be a candidate for investigation. This would more or less only be formally, as V1 works well enough. It might be, though, that V2 works just as well and is quicker. For such an investigation there would be two variants. In one $s_{add}$ would be added in steps over edges and corners alike, if the corresponding cell is occupied. In the other method one would multiply 1 or $\sqrt{2}$ with a parameter $s_{mult}$. 

In certain situations the influence of large jams progresses over long ranges, sometimes infinitely far. This was least pronounced for Chebyshev metric. While the movement of the agents at first seems to be unrealistic using Chebyshev metric, as the limited range of influence makes the agents try to evade the jam only much shorter in front of the jam, this could be used to simulate the influence of obstacles, smoke or darkness, where visibility range is limited as well. However, the trajectories generated by the influence of Chebyshev dynamic distance potential fields only create some associations of the behavior of real pedestrians in such situations. It might well be that actually there are large differences.

In \cite{Kretz2008c} it was shown that the error of V1 depends on the number of corners the potential has to flow around. Scenarios with a large number of such corners were not investigated in the present work. So it might be that for such geometries the V1 metric cannot be used as it might produce unrealistic results. This might be hard to prove, though, as the complexity of such a scenario a) makes it hard to develop an idea of how real pedestrians might walk and b) if this is possible, it might still be difficult to clearly detect deviations from this behavior in the simulation.

Whether one uses the method proposed in this contribution or not: a result of repeated simulation of pedestrian dynamics is the field $T_1(\vec{x},t)$ of travel times from some position to the destination. Just like the correction of the static distance potential field by the dynamic one, $T_1(\vec{x},t)$ is or at least should be a better estimation of the real travel times. Therefore one could use $T_1(\vec{x},t)$ similar to $S_{dyn}(\vec{x},t)$ in a second set of simulations to calculate a $T_2(\vec{x},t)$. The hope is that simulating iteratively converges: $|T_{i+1}(\vec{x},t)-T_{i}(\vec{x},t)|<\epsilon$ $\forall$ $i\geq j$ for a fixed $j$.  This would be a generalized dynamic assignment \cite{Friedrich2000} for fully two dimensional application, similar as in \cite{Hoogendoorn2004b}. For such an iterative strategy one would have to define an extrapolation method for the travel times of those locations that were not accessed by agents in the previous iteration. The importance and the requirements for the quality of this method are the higher the more the initial simulations deviate from the final solution. In this respect the method proposed in this work can be assumed to be very helpful.

\nocite{_TGF2001,_PED2001,_PED2003,_PED2005,_ACRI2006,_PED2008,_ACRI2008}
\def\newblock{\hskip .11em plus .33em minus .07em}
\bibliographystyle{utphys_quotecomma}
\bibliography{QuickestPath}

\end{document}